\documentstyle[12pt,epsfig]{article}
\def\beq{\begin{equation}}
\def\eeq{\end{equation}}
\def\bea{\begin{eqnarray}}
\def\eea{\end{eqnarray}}
\def\nnb{\nonumber}

\def\tY{\tilde Y}
\def\tal{\tilde \alpha}
\newcommand{\wti}{\widetilde}
\newcommand{\gsim}{\lower.7ex\hbox{$\;\stackrel{\textstyle>}{\sim}\;$}}
\newcommand{\lsim}{\lower.7ex\hbox{$\;\stackrel{\textstyle<}{\sim}\;$}}

\begin{document}
%\twocolumn[\hsize\textwidth\columnwidth\hsize\csname@twocolumnfalse\endcsname
%\begin{flushright}
%\baselineskip=12pt
%IC/2001/\\
%hep-ph/
%\end{flushright}
%\vskip 0.2cm

\begin{center}
\vspace{-3ex}{
                      \hfill hep-ph/0304130}\\[2mm]
{\Large\bf
%\title{
Neutrino Bilarge Mixing and Flavor Physics in the Flipped SU(5) Model}

\vspace{0.6cm}
%\author{
Chao-Shang Huang$^a$, Tianjun Li$^b$ and Wei Liao$^c$ \\
\vspace{0.3cm}
        $^a$ Institute of Theoretical Physics, Academia Sinica, P. O. Box 2735, \\
             Beijing 100080,  China\\
        $^b$ School of Natural Science, Institute for Advanced Study,  \\
             Einstein Drive, Princeton, NJ 08540, USA\\
        $^c$ The Abdus Salam International Center for Theoretical Physics,\\
              Strada Costiera 11, 34014 Trieste, Italy \\
%}
\end{center}
\begin{abstract}
%\abstract{
We have constructed a specific supersymmetric flipped SU(5) GUT
model in which bilarge neutrino mixing is incorporated. Because
the up-type and down-type quarks in the model are flipped in the
representations ten and five with respect to the usual SU(5), the
radiatively generated flavor mixing in squark mass matrices due to
the large neutrino mixing has a pattern different from those in
the conventional SU(5) and SO(10) supersymmetric GUTs. This leads
to phenomenological consequences quite different from SU(5) or
SO(10) supersymmetric GUT models. That is, it has almost no impact
on B physics. On the contrary, the model has effects in top and
charm physics as well as lepton physics. In particular, it gives
promising prediction on the mass difference, $\Delta M_D$, of the
$D-{\bar D}$ mixing which for some ranges of the parameter space
with large $\tan\beta$ can be at the order of $10^9 ~\hbar
~s^{-1}$, one order of magnitude smaller than the experimental
upper bound. In some regions of the parameter space $\Delta M_D$
can saturate the present bound. For these ranges of parameter
space, $t \to u,c+h^0$ can reach $10^{-5}-10^{-6}$ which would be
observed at the LHC and future $\gamma-\gamma$ colliders.
%}
\end{abstract}

%\pagebreak

%\begin{document}

\section{Introduction}\label{sec1}
In recent years great progresses have been made on the
flavor physics. Atmospheric neutrino~\cite{skatm} and solar neutrino~\cite{sno}
experiments together with the reactor neutrino~\cite{kamland,chooz}
experiments have established the
oscillation solution to the solar and atmospheric neutrino anomalies
~\cite{shiozawa,phs,also}. The solution tells us that neutrinos have masses
and mix with themselves in the propagations, {\it i.e.},
they oscillate~\cite{ponte,Maki:1962mu,msw}. The recent result from the
super-K collaboration and the combined analysis on the
solar neutrino experiment result and the KamLAND experiment result give
the best fit points~\cite{shiozawa,phs}
\bea
& \Delta m^2_{23}=2.5 \times 10^{-3} ~{\textrm eV}^2,
~~ \sin^2 2\theta_{23}=1.0;
\nnb \\
& \Delta m^2_{12}=7.3 \times 10^{-5} ~{\textrm eV}^2,
 ~~ \tan^2\theta_{12}=0.41;
\label{sol}
\eea
where $\Delta m^2_{ij}=m^2_{\nu_i}-m^2_{\nu_j}$
is the mass squared differences of the neutrinos in the mass
eigenstates (possible signs neglected here), and $\theta_{ij}$ are
the two-neutrino mixing angles. $\theta_{12}$ is for the solar
neutrino oscillation and $\theta_{23}$ is for the atmospheric
neutrino oscillation. Moreover, the CHOOZ experiment made a
constraint on the $\theta_{13}$ for the mass differences observed
in the atmospheric and solar neutrino experiments~\cite{chooz}
\bea
\label{chooz} |\sin\theta_{13}| \lsim 0.16.
\eea

Understanding these masses and mixings is a challenge. The
smallness of the masses can be understood via the see-saw
mechanism~\cite{ss}. Namely, the heavy right-handed
Majorana neutrinos, whose masses violate the lepton number symmetry,
simply decouple in the low energy physics and give very small
lepton number violating effects in the low energy phenomena,
which are the extremely small Majorana masses of the left-handed
neutrinos. The present experiments allow three typical solutions
to the neutrino masses: the degenerate case with $m_1 \sim
m_2 \sim m_3 \sim 10^{-1}$eV; hierarchy case with $m_1 \ll m_2
\approx \sqrt{\Delta m^2_{21}}$, $m_2 \ll m_3 \approx \sqrt{\Delta m^2_{32}}$;
inverse hierarchy case with $m_3 \ll m_1 \sim m_2 \sim
\sqrt{\Delta m^2_{23}}$.

In view of the beautiful picture raised by
Weinberg-Wilczek-Zee-Fritsch (WWZF)~\cite{wwzf} that the small
quark mixing in the CKM matrix is related to the large quark mass
hierarchy, people feel challenged a lot by the presence of the
neutrino bilarge mixing. Some people worried that we should have
degenerate mass matrix to understand it. However, in the WWZF
scenario only the symmetric mass matrix is used. If allowing
asymmetric form for the mass matrix which for example may well be
generated by the elegant Froggatt-Nielsen (FN)
mechanism~\cite{FN}, we can accomodate the large mass hierarchy
with large mixings~\cite{asy}. If one works with an effective
theory, {\it e.g.}, the minimal supersymmetric Standard Model
together with right-hand neutrinos (MSSM+N), at a low energy scale
(say, the electro-weak scale), one can content oneself by using
the WWZF scenario to understand the smallness of quark mixing, and
the see-saw and FN mechanism to understand the largeness of
neutrino mixing. However, if one works with a theory, {\it e.g.},
a grand unification theory (GUT), in which quarks and leptons are
in a GUT multiplet, one has to answer: can we explain
simultaneously the smallness of quark mixing and the largeness of
neutrino mixing in the theory? If we can, then what are the
phenomenological consequences in the theory? There are several
recent works to tackle these problems in SU(5) or SO(10) GUTs
~\cite{bdv,lfv3,cmm,mvv,bsv,hs}.

As known for a long time in the framework of the supersymmetric
see-saw mechanism, we are able to predict the lepton flavor
violating (LFV) effects~\cite{lfv,lfv2}. The flavor structure in
the neutrino Yukawa couplings can be transformed to the soft SUSY
breaking masses of the left-handed sleptons which then give
implications on the $\mu \to e \gamma$ and $\tau \to \mu,e+\cdots$
processes. In the models of the grand unification theory, {\it
e.g.}, in $SU(5)$ or $SO(10)$, quark fields are unified with
leptons in the representations of GUT gauge group, and from the
lepton Yukawa couplings we are able to get the flavor mixings in
down-type squark mass matrix~\cite{hkr,bdv,cmm,hs}. These flavor
mixings are something beyond those described by the CKM mixing and
can give the interesting phenomenological implications for the
Kaon and B meson physics. This approach has recently been used in
SUSY SU(5) and SO(10) GUTs which incorporate the bilarge neutrino
mixing and can give the significant phenomenological predictions
if the bilarge neutrino mixing is from the lepton Yukawa
couplings~\cite{bdv,cmm,hs}.

In the present paper, we address the problems in the
supersymmetric flipped $SU(5)$ model.  As we know, the flipped
$SU(5)$ unification model has several advantages: (i) $SU(5)\times
U(1)$ is the minimal unified gauge group which provides neutrino
masses; (ii) without high dimension Higgs representations; (iii)
the natural splitting of the doublet and triplet components of the
Higgs pentaplets and consequently the natural avoidance of
dangerous dimension-5 proton decay operators; and (iv) the natural
appearance of a see-saw mechanism for neutrino masses.
Furthermore, in the context of 4-dimensional free fermionic string
model-building, at level one ($k=1$) of the Kac-Moody
algebras~\cite{eln}, one can only obtain the Standard-like
Model~\cite{AFDVN}, the Pati-Salam Model~\cite{IAGLJR} and the
flipped $SU(5)$ model~\cite{JLLDVN},
 because the dimensions of Higgs fields in the spectra
are smaller than that of the adjoint representation. Of these,
only flipped $SU(5)$ actually unifies the non-abelian gauge groups
of the Standard Model (SM). And a proliferation of $U(1)$ factor
is the norm.

We show that the fermion mass hierarchies and the mixings of quarks
and leptons can all be well accommodated in the flipped $SU(5)$
model. Moreover, because in flipped $SU(5)$ the up-type and down-typ
 quarks
are flipped in the representations ten and five with respect to
the usual SU(5), the one loop radiative corrections to sfermion masses
have a pattern different from those in SU(5) or SO(10) and
consequently different phenomenological implications. We find that
in flipped $SU(5)$ new effects appear in top and charm physics, in
addition to lepton physics, and no effects in B physics, which is
a novel feature quite different from the usual SU(5) and SO(10)
GUTs and can be tested by incoming experiments.

The paper is organized as follows. In section II, after a brief
review of the flipped $SU(5)$ model, we construct a specific model
in which the bilarge mixing can be accomodated in the Yukawa
couplings as well as the fermion mass hierarchy and the quark
mixing. We will then discuss the radiatively corrected SUSY
breaking soft terms with universal SUSY breaking at the Planck scale.
In section III, we discuss the phenomenological consequences of
our model. Section IV is devoted to conclusions and discussions.
Finally, some conventions in MSSM+N are given in Appendix A, and
the renormalization group equations (RGEs) in MSSM+N and flipped
SU(5) are given in the Appendices B and C, respectively.
Throughout the paper we assume no $CP$ violating phases appearing
the leptonic sector to simplify the discussion.

\section{Flipped SU(5) Model and the Flavor Structure}\label{sec2}
\subsection{The Flipped $SU(5)$ Model}\label{sec2.1}
In this subsection, we would like to briefly review the Flipped
$SU(5)$~\cite{smbarr, dimitri} and construct a specific flipped
SU(5) model. The gauge group for flipped $SU(5)$ model is
$SU(5)\times U(1)_{X}$, which can be embedded in $SO(10)$ model.
We can define the generator $U(1)_{Y'}$ in $SU(5)$ as 
\bea 
T_{\rm U(1)_{Y'}}={\rm diag} \left(-{1\over 3}, -{1\over 3}, -{1\over 3},
 {1\over 2},  {1\over 2} \right).
\label{u1yp}
\eea
The hypercharge is given by
\bea
Q_{Y} = {1\over 5} \left( Q_{X}-Q_{Y'} \right).
\label{ycharge}
\eea

There are three families of fermions with the following
$SU(5)\times U(1)_{X}$ transformation proporties
\bea
F_i=(10, 1),~ {\bar f}_i=(\bar 5, -3),~l_i^c=(1, 5),
\label{smfermions}
\eea
where $i=1, 2, 3$.
As an example, the particle assignments for the first family are
\bea
F_1=(Q_1, D^c_1, N_1),~{\bar f}_1=(U^c_1, L_1),~l_1^c=E^c_1.
\label{smparticles}
\eea
$Q$ and $L$ are the superfields of the
quark and lepton doublets, $U^c$, $D^c$, $E^c$ and $N$ are the
$CP$ conjugated superfields for the right-handed up-type quark,
down-type quark, lepton and neutrino.

To break the GUT and electroweak symmetries, we introduce two pairs
of Higgs representations
\bea
H=(10, 1),~{\bar H}=({\bar {10}}, -1),~h=(5, -2),~{\bar h}=({\bar {5}}, +2).
\label{Higgse1}
\eea
We label the states in the $H$ multiplet by the same symbols as in
the $F$ multiplet, and for $\bar H$ we just add ``bar'' above the fields.
Explicitly, the Higgs particles are
\bea
H=(Q_H, D_H^c, N_H),~{\bar H}= ({\bar {Q}}_{\bar H}, {\bar {D}}_{\bar H}^c, {\bar {N}}_{\bar H}),
\label{Higgse2}
\eea
\bea
h=(D_h, D_h, D_h, H_1),~{\bar h}=({\bar {D}}_{\bar h}, {\bar {D}}_{\bar h},
{\bar {D}}_{\bar h}, H_2).
\label{Higgse3}
\eea
We also add one singlet $S$.

To break the $SU(5)\times U(1)_{X}$ gauge symmetry down to the SM
gauge symmetry, we introduce the GUT superpotential 
\bea
{\it W}_{\rm GUT}=\lambda_1 H H h + \lambda_2 {\bar H} {\bar H} {\bar
h} + S ({\bar H} H-M_{\rm V}^2). 
\label{spgut} 
\eea 
There is only
one F-flat and D-flat direction, which can always be rotated along
the $N_H$ and ${\bar N}_{\bar H}$ directions. So, we obtain that
$<N_H>=<{\bar N}_{\bar H}>=M_{\rm V}$. In addition, the
superfields $H$ and $\bar H$ are eaten or acquire large masses via
the supersymmetric Higgs mechanism, except for $D_H^c$ and ${\bar
D}_{\bar H}^c$. And the superpotential $\lambda_1 H H h$ and
$\lambda_2 {\bar H} {\bar H} {\bar h}$ combine the $D_H^c$ and
${\bar D}_{\bar H}^c$ with the $D_h$ and ${\bar {D}}_{\bar h}$,
respectively, to form the massive eigenstates with masses
$\lambda_1 <N_H>$ and $\lambda_2 <{\bar N}_{\bar H}>$. So, we
naturally have the doublet-triplet splitting due to the missing
partner mechanism. Because the triplets in $h$ and ${\bar h}$ only have
small mixing through the $\mu$ term, the higgsino-exchange mediated
proton decay are negligible, {\it i.e.},
we do not have dimension-5 proton
decay problem. The singlet $S$ can have GUT scale mass and decouple
in the low energy theory.

The SM fermion masses are from the following
superpotential\footnote{In popular flipped SU(5) models one
usually introduces more singlets and constructs a renormalizable
superpotential. Here instead we introduce a non-renormalizable
term to give the right-hand neutrino masses in order to avoid
introducing more fields.}
\bea 
{\it W}_{\rm Yukawa} = {1\over 8}
F_i (Y_{10})_{ij} F_j h + F_i (Y_{\bar 5})_{ij} {\bar f}_j {\bar
h}+ l_i^c (Y_1)_{ij} {\bar f}_j h + \mu h {\bar h}
+{1\over {2M_*}} F_i (Y_R)_{ij}
F_j {\bar H} {\bar H}. \label{potgut}
\eea
After the $SU(5)\times
U(1)_X$ symmetry is broken down to the SM group, the
superpotential gives 
\bea 
{\it W_{SSM}}&=&
D^c_i(Y_{10})_{ij}Q_jH_1+U^c_i (Y_{\bar 5})_{ji}Q_j H_2
+E^c_i(Y_1)_iL_jH_1+N_i(Y_{\bar 5})_{ij} L_j H_2 \nnb \\
&& + \mu H_1 H_2+ \frac{1}{2} (M_N)_{ij} N_i N_j
 + \cdots (\textrm{decoupled below $M_{GUT}$}). 
\label{poten1}
\eea
Thus, at the $M_{GUT}$ scale we have
\bea 
Y_U=Y^T_{\bar 5},
~~ Y_D=Y_{10}, ~~ Y_N=Y_{\bar 5},~~ Y_E=Y_1.
\label{bound1}
\eea
The right-handed neutrino mass matrix $M_N$ is given by $Y_R$.

Assuming the right-handed neutrinos are heavy, they decouple at
the low energy scale. After we integrate out the right-handed neutrinos,
the left-handed neutrino Majorana masses are given by
\bea 
{\it W}_{\rm m_{\nu_L}} = {{M_*}\over {M_V^2}} (Y^T_{\bar 5} Y_R^{-1}
Y_{\bar 5})_{ij} L_i L_j H_2 H_2~.~
\label{numajma}
\eea
 Thus, if
$(Y_R)_{33} M_V^2/M_* \sim 10^{14}$ GeV, we obtain the correct
$\tau$ neutrino ($\nu_{\tau}$) mass implied by the atmospheric
neutrino oscillation experiment because the $\tau$ neutrino
($\nu_{\tau}$) Dirac mass is equal to the top quark mass at the
GUT scale due to Eq. (\ref{bound1}). In addition, the left-handed
neutrino Majorana mass matrix, which is symmetric, can be taken to
be arbitrary because the Majorana mass matrix $Y_R$
 for the right-handed
neutrinos is arbitrary.

We can see from Eq. (\ref{potgut}) that the up-type quark Yukawa
matrix is the transpose of the neutrino Dirac Yukawa matrix and
the down-type quark Yukawa matrix is symmetric. Because in the
superpotenyial $W_{SSM}$, the up-type quark mass matrix, the
lepton mass matrix and the symmetric left-handed neutrino Majorana
mass matrix are arbitrary, we can generate the correct SM fermion
mass matrices, the CKM matrix and the neutrino mixing matrix,
although the down-type quark mass matrix is symmetric.

To be concrete, we present a realistic sample for the SM fermion
mass matrices in the Flipped $SU(5)$ model which accomodates the
bilarge neutrino mixing. The discussion will be made without
considering the CP violating phases, as noted in Introduction.

The ratios of the masses at the GUT scale
\cite{Fusaoka:1998vc}
are
\begin{eqnarray}
 m_u : m_c : m_t \sim \lambda^7 : \lambda^4 : 1 \ ,
\label{uptmass}
\end{eqnarray}
\begin{eqnarray}
 m_d : m_s : m_b \sim \lambda^4 : \lambda^2 : 1 \ ,
\end{eqnarray}
\begin{eqnarray}
 m_e : m_\mu : m_\tau \sim \lambda^5 : \lambda^2 : 1 \ .
\end{eqnarray}
where $\lambda=0.22$.
And the CKM matrix is given by
\begin{eqnarray}
 V_{\rm CKM} \sim \left(
\begin{array}{ccc}
 1 & \lambda & \lambda^3 \\
 -\lambda& 1 & \lambda^2 \\
 -\lambda^3& -\lambda^2 & 1\\
\end{array}
\right)~.~\,
\label{ckm}
\end{eqnarray}

In the hierarchy case, we can estimate the ratio of
the neutrino masses
\bea
 m_{\nu_2} : m_{\nu_3} \sim
\sqrt{\frac{\Delta m^2_{\rm solar}}{\Delta m^2_{\rm atm.}}}
\sim \epsilon' : 1\ ,
\label{neutrino-ratio}
\eea
where $\epsilon'$ is about $0.2$ according to Eq. (\ref{sol}).
The neutrino mixing matrix
can be parametrized in the standard way
\bea
U &=&
\pmatrix{c_{12}c_{13}& s_{12}c_{13} &  s_{13}e^{-i\delta_{13}} \cr
   -s_{12}c_{23} -c_{12}s_{23}s_{13} e^{i\delta_{13}}  &
   c_{12}c_{23} - s_{12}s_{23}s_{13} e^{i\delta_{13}} & s_{23} c_{13} \cr
   s_{12}s_{23} - c_{12}c_{23} s_{13} e^{i\delta_{13}}  &
   -c_{12} s_{23} - s_{12}c_{23} s_{13} e^{i\delta_{13}}& c_{23}c_{13} },
\label{stpara}
\eea
where $c_{ij}=\cos\theta_{ij}$, $s_{ij}=\sin\theta_{ij}$, and
 $\delta_{13}$
is the $CP$ violating phase which is taken as zero as those
Majorana phases in our discussion, as we noted in the
Introduction. In view of Eq. (\ref{chooz}), one has $c_{13}\gsim
0.98$, then, to a good approximation, one has
\bea
 U &=&
\pmatrix{c_{12}& s_{12} &  s_{13} \cr
   -\frac{s_{12}}{\sqrt{2}}   &
   \frac{c_{12}}{\sqrt{2}}  & \frac{1}{\sqrt{2}}  \cr
   \frac{s_{12}}{\sqrt{2}} &
   -\frac{c_{12}}{\sqrt{2}} & \frac{1}{\sqrt{2}} },
\label{stpara1}
\eea
 where $c_{23}=s_{23}=\frac{1}{\sqrt{2}}$,
$c_{13}=1$, and $s_{12} < c_{12}$ with $\theta_{12}\sim 33^0$ are
globally consistent with all neutrino experiments known so far.

The up-type quark Yukawa and the neutrino Dirac Yukawa couplings can take
the following form
\begin{eqnarray}
Y_U = (Y_{\nu})^T \sim y_t
\left(
\begin{array}{ccc}
 \lambda^7 & \lambda^6 & \lambda^4 \\
 \lambda^5 & \lambda^4 & \lambda^2 \\
 \lambda^3 & \lambda^2 & 1\\
\end{array}
\right)\ ,\ \
\end{eqnarray}
the down-type Yukawa coupling $Y_D$ takes the symmetric form
\begin{eqnarray}
Y_D \sim y_b
\left(
\begin{array}{ccc}
 \lambda^4 & 0 & 0 \\
 0 & \lambda^2 & 0 \\
 0 & 0 & 1\\
\end{array}
\right)\ ,\ \
\end{eqnarray}
and the lepton Yukawa coupling $Y_E$ can take the form
\begin{eqnarray}
Y_E \sim \frac{y_\tau}{\sqrt{2}}
\left(
\begin{array}{ccc}
 \sqrt{2} a \lambda^5 &  \sqrt{2} b \lambda^5 & \sqrt{2} \epsilon \lambda^5 \\
-b \lambda^2 & a \lambda^2 & \lambda^2 \\
b & -a & 1 \\
\end{array}
\right)\ ,\ \
\end{eqnarray}
where $a^2+b^2=1$, $\epsilon$ is smaller than one.  In the base
where the charged leptons are mass eigenstates and assuming that
the same unitary matrix $U$ makes both the lepton mass matrix and
the left-handed neutrino Majorana mass matrix diagonal, we obtain
\bea
\sin\theta_{13} \approx \epsilon ,~~ \tan\theta_{12} \approx
\frac{b}{a}, ~~ \tan\theta_{23} =1.
\label{theta}
\eea
The bilarge
neutrino mixing, Eq. (\ref{stpara1}), follows if $\epsilon\lsim
0.16$, $b$ and $a$ are both of the order one and $b < a$. The
prediction on $\theta_{13}$ heavily depends on the
non-diagonal entries in the left-handed neutrino Majorana mass
matrix since we would like to get a small $\theta_{13}$. Because
the right-handed neutrino mass matrix is arbitrary, as noted above,
an appropriate form of the right-handed neutrino mass matrix can
be taken to maintain the relation of $\theta_{13}$ in Eq.
(\ref{theta}). The left-handed neutrino Majorana mass matrix, Eq.
(\ref{numajma}), is given for the flavor fields $\nu_{\alpha}$,
$\alpha=e,\mu,\tau$. By the unitary transformation $U$ given in
Eq. (\ref{stpara1}), we obtain the left-handed neutrino masses
which can have mass hierarchy as
\bea
m_{\nu_1}:m_{\nu_2}:m_{\nu_3} \sim \epsilon'' : \epsilon' :1,
\eea
where the mass eigenvalues $m_{\nu_i}$ and consequently
$\epsilon''$ and $\epsilon'$ are determined by the matrix
($Y^T_{\bar 5} Y_R^{-1} Y_{\bar 5}$) ( see Eq.(\ref{numajma}) )
with $Y^T_{\bar 5}=Y_U$. In order to get $\epsilon'' <
\epsilon'\sim \lambda$, the eigenvalues, $M_{N_i}$, of the
right-handed neutrino
Majorana mass matrix $M_N$ should have a hierarchy as large as
\bea
M_{N_2} : M_{N_3} \sim \lambda^7 :1
\eea
which for the
$M_{N_3} \sim 10^{14-15}$ GeV gives $M_{N_2} \sim 10^{9-10}$
GeV.

In the above discussions, the hierarchies in fermion mass spectra
are obtained due to the hierarchies in Yukawa couplings.
An elegant mechanism to understand this fermion mass hierarchy problem is
the Froggatt-Nielsen mechanism~\cite{FN}.
However, the $U(1)$ global symmetry is not enough to generate the correct
fermion masses, CKM matrix and neutrino mixing matrix.
 So, we need to consider the larger global symmetry,
for instance, $SU(2)\times U(1)$ family symmetry. This is out of
the scope of this paper and we will not consider it here.

Before proceeding, a remark is in place. The presence of the
non-renormalizable operators might affect the predictions based on
the renormalizable Yukawa couplings in Eq. (\ref{potgut}).
In our model, we do not have the dimension-5 non-renormalizable operators
which can give the SM fermion masses except the operators which give
the right-handed neutrino Majorana masses.
There are dimension-6 non-renormalizable operators,
for example,
\bea
{\it W}_{\rm np} =
{1\over M_*^2} \lambda^3_{ij} [F_i ({\bar H} h)] F_j H + {1\over
M_*^2} \lambda^4_{ij} [F_i {\bar H} ] [ {\bar f}_j (H {\bar h})],
\label{nrsp}
\eea
which can give the down-type
quark masses $\lambda^3_{ij} {{v_1 <N_H> <{\bar N}_{\bar
H}>}\over\displaystyle M_*^2} {\bar d}_{Rj} d_{Li}$ and the
neutrino Dirac masses $\lambda^4_{ij} {{v_2 <N_H>
<{\bar N}_{\bar H}>}\over\displaystyle M_*^2} N_i \nu_j$,
respectively.
 However, the effects from the
dimension-6 non-renormalizable operators can be safely neglected
if taking the cutoff scale of the theory high enough. For
instance, taking $M_*=M_{Pl}$ implies that their effects are
suppressed by $M^2_{GUT}/M^2_{Pl} \approx 10^{-4}$. So, the
predictions described in Eq. (\ref{theta}) are not affected by
these operators~\footnote{ The dimension-5 operators can have
larger effects. We have used one to generate the right-handed
neutrino masses. More operators, like $(F_i {\bar f}_{\bar h}) F_j
H/M_*$ and $(F_i {\bar H}) ({\bar f}_j f_h)/M_*$ can appear if we
introduce one pair of Higgs $f_h=(5, 3)$ and ${\bar f}_{\bar h}=
(\bar 5, -3)$ under $SU(5)\times U(1)_{X}$. However, one may
reintroduce the dimension-5 proton decay operators
 via the higgsino exchange. So, we are not
going to consider this possibility.}. Although there are also RGE
effects on the mixing matrix, as observed for the normal
hierarchical case, the RGE effects are mild~\cite{rgeff}. In this
paper, we will therefore use the mixing matrices established in
the low energy phenomena above the see-saw scale.

We conclude for this subsection that the
 bilarge neutrino mixing can be well
accomodated in the lepton Yukawa couplings in the flipped SU(5)
model. The point is that
for flipped $SU(5)$ model, the up-type quark mass matrix and the
lepton mass matrix are arbitrary, and the symetric Majorana mass matrix
for left-handed neutrino is arbitrary after the right-handed neutrinos decouple.
Therefore, although the mass matrix for
down-type quark is symmetric, we do have enough degrees of freedoms to
produce the correct
GUT scale SM fermion masses, CKM matrix and neutrino mixing matrix.

\subsection{One Loop Radiative Corrections of Sfermion Masses}\label{sec2.2}

In order to establish notations and see where the difference of
phenomenological consequences between the flipped and conventional
SU(5) comes from, let us first illustrate a little on how the
flavor changing terms arise in the supersymmetric theories. In
this and next subsections, we shall use some useful conventions and
definitions in the Minimal Supersymmetric Standard Model (MSSM)
plus the right-handed neutrino fields (MSSM+N) given in Appendix
A. In MSSM+N, the superpotential can be written as follows
 \bea 
{\it W}_{SSM} &=& D^c_i(Y_D)_{ij}Q_jH_1+U^c_i
(Y_U)_{ij}Q_j H_2 +E^c_i(Y_E)_{ij}L_jH_1 \nnb \\
&&+N_i(Y_N)_{ij}L_jH_2 + \mu H_1 H_2 +\frac{1}{2}(M_N)_{ij}N_iN_j.
\label{spssm}
\eea
$H_2$ and $H_1$ are the
Higgs doublets which give the Dirac masses to the up-type quark
(neutrino) and down-type quark (lepton), respectively. Assuming
the universal SUSY breaking at high energy scale, we can get the
radiative corrections to the mass of squark doublets, $m^2_{Q}$
(see Appendix A for the convention), with the following
form~\cite{fla}
\bea
 \delta m^2_Q &\propto & Y^\dagger_U Y_U
+Y^\dagger_D Y_D \nnb.
\eea
This just tells us that two couplings
of $Q$ in the Eq. (\ref{spssm}) both contribute. After the
electroweak (EW) symmetry breaking, $m^2_Q$ gives $m^2_{U_L}$ and
$m^2_{D_L}$. CKM mixing in the charged current interaction is
obtained after diagonalising $Y_U$ and $Y_D$. Typically for the
left-handed quark fields, the diagonalization is concerning
$Y^\dagger_U Y_U$ and $Y^\dagger_D Y_D$. As can be seen clearly we
will have misalignment between the quark and the squark mass
matrix. Transforming for example the down-type squark fields
simultaneously with the down-type quarks can diagonalise
$Y^\dagger_D Y_D$ but can not diagonalise $m^2_Q$. Similarly for
the up-type squarks. Thus, we have
\bea
\delta ({\wti m}^2_{U_L})_{ij}
\propto (K y^2_D K^\dagger)_{ij}
\approx K_{i3} y^2_b K^*_{j3} \nnb \\
\delta ({\wti m}^2_{D_L})_{ij} \propto (K^\dagger y^2_U K)_{ij}
\approx K^*_{3i} y^2_t K_{3j}.
\label{flamssm}
\eea
where $y_U$
and $y_D$ are the diagonalised Yukawa couplings and $K$ is the CKM
matrix. ${\wti m}^2$ is the mass squared written in the super-CKM
base in which $Y_{U,D,E}$ are all diagonalised and the sfermion
fields are transformed simultaneously with the fermion fields (see
Appendix A). In the super-CKM base, we can see directly the
 extra flavor structures
in the off-diagonal terms of these soft SUSY breaking masses.
 One of the main features of these corrections is that
they are proportional to the corresponding flavor changing CKM
matrix element. Typically $\delta({\wti m}^2_{U_L})_{ij}$ is given
by the third column of the CKM matrix and $\delta({\wti
m}^2_{D_L})_{ij}$ is given by the third row.

Similar story happens in the flipped $SU(5)$ model. As can be
seen in Eqs. (\ref{potgut}) and (\ref{bound1}), $Y_U$ and $Y_E$
come seperately from the Yukawa couplings of five representation
to the ten representation and of the five to the singlet
representation. Since these two couplings both give the SM fermion
Yukawa couplings, after transforming to the super-CKM base we can similarly
get $\delta({\wti m}^2_{U_R})_{ij}$ and $\delta({\wti m}^2_{E_L})_{ij}$
radiatively corrected by the different elements of the same matrix.
The story is different with the conventional
$SU(5)$ embedding. In the conventional $SU(5)$ theory
$10=(Q,U^c,E^c)$, ${\bar 5}=(D^c,L)$ and $1=N$. $Y_D$ and $Y_E$
both come from the Yukawa couplings of five representation to the
ten representation. We have $\delta({\wti m}^2_{D_R})_{ij}$ and
$\delta({\wti m}^2_{E_L})_{ij}$ being radiatively corrected by the
coupling of five to the singlet representation which is basically
the Yukawa coupling of the right-handed neutrino. Consequently
$\delta({\wti m}^2_{D_R})_{ij}$ and $\delta({\wti m}^2_{E_L})_{ij}$
have similar dependences on the same matrix. Thus, what happens
in the flipped $SU(5)$ is not quite similar to what happens in
the conventional $SU(5)$ theory. And we have new phenomena in
the charm and top quark physics, which will be discussed in detail
in the next subsection and section III.

\subsection{Low Energy Implications}\label{sec2.3}

We can define the SUSY breaking soft terms for the multiplets of
the flipped $SU(5)$ model as follows
\bea 
- \Delta {\cal L} &=&
\wti{F}_i^* (m^2_{10})_{ij}\wti{F}_j +{\wti {\bar f}}_i^*
(m^2_{\bar 5})_{ij} {\wti {\bar f}}_j +{\tilde l}_i^{c*}
(m^2_1)_{ij} {\tilde l}_j^c + m_h^2 {\wti h}^* {\wti h}+m_{\bar
h}^2 {\wti {\bar h}^*}
{\wti {\bar h}} \nnb \\
&& +\bigg[\frac{1}{8} \wti{F}_i(Y^A_{10})_{ij}\wti{F}_j {\wti h} +\wti{F}_i
(Y^A_{\bar 5})_{ij}\wti{\bar f}_j \wti{\bar h} +{\tilde
l}_i^c(Y^A_1)_i{\wti {\bar f}}_j{\wti h} +\mu B{\wti h} \wti{\bar
h} \nnb \\
&& +\frac{1}{2} M_5 \lambda_5 \lambda_5
+\frac{1}{2}M_X \lambda_X \lambda_X+h.c.].
\label{softgut}
\eea
$\lambda_5$ and $\lambda_X$ are respectively the gaugino fields of the
$SU(5)$ and $U(1)_X$ gauge groups. $M_5$ and $M_X$
are the gaugino masses.
At $M_{GUT}$ scale, the
matching conditions are 
\bea 
& Y^A_U=(Y^A_{\bar 5})^T, ~~
Y^A_D=Y^A_{10},
~~ Y^A_N=Y^A_{\bar 5},~~ Y^A_E=Y^A_1 \nnb \\
& m^2_Q=m^2_{10}, ~~ m^2_D=(m^2_{10})^T, ~~ m^2_U=(m^2_{\bar 5})^T,
~~ m^2_L=m^2_{\bar 5}, ~~m^2_E=(m^2_1)^T.
\label{bound2}
\eea
${\wti h}$ et.al refer the scalar part of the corresponding
superfields.

Similar to that in the MSSM, we do have $m^2_{\bar 5}$ corrected
by the presence of two Yukawa couplings, $Y_5$ and $Y_1$:
\bea
\delta m^2_{\bar 5} \propto c_1 Y_{\bar 5}^\dagger Y_{\bar 5}
+c_2 Y_1^\dagger Y_1
= c_1 (Y_U Y_U^\dagger)^T +c_2 Y_E^\dagger Y_E.
\eea
$c_{1,2}$ are coefficients.
Again similar to what happend in the MSSM, $Y_{\bar 5}$ and $Y_1$ give
masses to the different SM fermions, {\it i.e.}, the up-type quarks
and leptons. The diagonalisations are concerning $Y_U Y_U^\dagger$
for the right-handed up-type quark and $Y_E^\dagger Y_E$
for the left-handed lepton fields. The resulting extra
flavor structures in the sfermion mass matrices of right-handed
up-type squarks and sleptons depend on the different
elements of the same matrix. The results are detailed in the
Appendix C.  According to them, we can get
the mass insertion parameters for the corresponding off-diagonal
mass terms. Mass insertion parameter $\delta_{ij}$ is defined
as $\delta_{ij}=M^2_{ij}/M^2_{\tilde f}$, where $M^2_{\tilde f}$
is the averaged mass squared of the $i$ and $j$ diagonal entries
of the mass matrix of ${\tilde f}$.
As an example, we have
\bea
(\delta^U_{RR})_{12} &\approx& -0.01 \frac{3 m_0^2+|A_0|^2}
{|M_{1 \over 2}|^2 (1+0.16 m_0^2/|M_{1 \over 2}|^2)}
(U_N)_{13}^* y_\tau^2 (U_N)_{23}, \label{flavor5}\\
(\delta^E_{LL})_{12} &\approx& -0.25\frac{3 m_0^2+|A_0|^2}
{0.6 |M_{1 \over 2}|^2+m_0^2} (U_N)_{31}^* y_t^2 (U_N)_{32}(1+\frac{1}{40}
ln\frac{M_{GUT}^2}{M_N^2}). \label{flavor6}
\eea
In the expression, we have approximated $({\wti m}_{f_{LL,RR}}^2)_{ii}
\approx 6 |M_{1 \over 2}|^2 + m_0^2$ ($i=1,2$ and $f=U$)
and $({\wti m}_{E_{LL,RR}}^2)_{jj}
\approx 0.6 |M_{1 \over 2}|^2+m_0^2$. D-term contributions after
the EW symmetry breaking have been neglected in the estimate.
We have also taken
$ln\frac{M_*^2}{M_{GUT}^2}=10$ for the estimate which means
that $M_* \doteq 3 \times 10^{18} \textrm{GeV} \approx M_{Pl}$
for $M_{GUT} \doteq 2.0 \times 10^{16}$GeV.
For $\delta^E_{LL}$
we have included the contributions
from the RGE running between the
$M_{GUT}$ scale and the $M_N$ scale shown in the Appendix B.
We see that the corrections to $\delta^E_{LL}$ can be quite large
if the relevant entries in $U_N$ can be of order one.

We can see clearly that
typically $\delta^E_{LL}$ probes the third column of the matrix $U_N$,
and $\delta^U_{LL}$ probes the third row of $U_N$. Similarly for
$\delta^E_{LR}$ and $\delta^U_{LR}$. Notice that to have sizeable
effects for the $\delta^U_{RR}$, we need large $\tan\beta$ because
it is corrected by the presence of lepton Yukawa couplings.
We can classify three typical scenarios by noticing that
$U_N=V_{N_L}^\dagger V_{E_L}$:\\
\\
~~~(i) Neutrino mixings are all from $U_N$ and we can take
$U= U_N^\dagger$ (see Eq. (\ref{neum}) for comparison).
\\
\\
~~~(ii) Only the atmospheric neutrino mixing comes from $U_N$ and
off-diagonal entries of
the first row and first column of $U_N$ are all small and negligible;\\
\\
~~~(iii) No large off-diagonal entries in $U_N$ and large mixing angles
come from $V_{N_R}$ and $M_N$.
\\

The first possibility is supported
by our discussions in the last subsection which can give the
interesting phenomenological predictions. We will concentrate
on the first case and comment on the second and the third one in the
last section of the paper.

For scenario (i), we have
\bea
&& (\delta^U_{RR})_{ij} \propto U_{3i} U_{3j}^*, ~~~ (\delta^U_{LR})_{3k}
\propto U_{33} U_{3k}^*,\\
&& (\delta^E_{LL})_{ij} \propto U_{i3} U_{j3}^*, ~~~ (\delta^E_{LR})_{k3}
\propto U_{k3} U_{33}^*, ~~~~\textrm{$i\neq j$, $k\neq 3$}.
\eea
As expected, the corrections are concerning the right-handed
squark and the left-handed sleptons. As we know, elements
in the third row of the matrix $U$ are of the same order
because of the bilarge mixing, then we have
\bea
|(\delta^U_{RR})_{12}| ~\sim~ |(\delta^U_{RR})_{13}|
~\sim~ |(\delta^U_{RR})_{23}|.
\label{rela}
\eea
In addition, in the case of large $\tan\beta$ for which
$y_\tau$ and $y_t$ are of comparable magnitudes
we have the following features for the moderate
vaules of $M_{1 \over 2}$, $m_0$ and $A_0$:\\
\\
(a)$(\delta^U_{RR})_{ij}$ is of order $10^{-2}$ and
$(\delta^E_{LL})_{23}$ is of the order of $10^{-1}$;\\
\\
(b) $(\delta^E_{LR})_{23}$ is further suppressed by the
VEV, $v_1=v \cos\beta$, and is of the order $10^{-2}-10^{-3}$
depending on the sfermion mass spectrum and is of
the same order as $(\delta^U_{LR})_{3k}$;\\
\\
(c) $(\delta^E_{LL})_{13,12}$ and $(\delta^E_{LR})_{13}$ are
further suppressed by the presence of the small magnitude of
$U_{13}$ (or $U_{e3}$).

\section{Phenomenology}
\subsection{Lepton Flavor Violation}\label{sec3.1}
Satisfying the precision $\mu \to e \gamma$ constraint
\bea
 Br(\mu \to e \gamma) < 1.2 \times 10^{-11},
\eea
puts order $10^{-3}$
upper bound on $(\delta^E_{LL})_{12}$ and order $10^{-5}$ upper
bound on $(\delta^E_{LR})_{12}$~\cite{ggms,mpr}. For moderate
values of $M_{1 \over 2}$, $m_0$ and $A_0$, they correspond to
having $|U_{e3}| \lsim 10^{-3}$. For the scenario discussed in the
section \ref{sec2.1}, they correspond to having $\epsilon \lsim
10^{-3}$. We can simply satisfy this constraint by assuming
$U_{e3}=s_{13}=0$ in the matrix $U$. Similar thing happens to
$\tau \to e + \cdots$ processes.

Since the $(\delta^E_{LL})_{23}$ is at the order of $10^{-1}$, we
are able to get the promising prediction on $\tau \to \mu \gamma$
through the $\tan\beta$ enhanced contributions~\cite{lfv3,mvv}. A
rough estimate in the mass insertion~\cite{mvv} shows that 
\bea
Br(\tau \to \mu \gamma) \approx 4. \times 10^{-7} \times \bigg(
\frac{1000 ~{\textrm GeV}}{m_{\tilde l}} \bigg)^4 \times
\bigg(\frac{\tan\beta}{50} \bigg)^2, 
\label{approlep}
\eea 
for $m_0= |A_0|=2 |M_{1
\over 2}|$. The prediction is close to the present $1.1 \times
10^{-6}$ upper bound~\cite{pdg}. In the formulae, $m_{\tilde l}$
is the average mass of the $\tau$ and $\mu$ sleptons. And we have
used $M_{N_3} = 5.4 \times 10^{14}$ GeV to get $m_{\nu_3}=0.05$
eV. If taking for example $|A_0|= 6 |M_{1 \over 2}|$ and $m_0 =2
|M_{1 \over 2}|$ (see Eq. (\ref{flavor5})) which are in the region
allowed by the constraints from the relic density of cold dark
matter as well as $(g-2)_{\mu}$ and $b\rightarrow s
\gamma$~\cite{cdm}, the $(\delta^E_{LL})_{23}$ can be at the order
of one. This will make Br($\tau \to \mu \gamma$) beyond the
experimental upper bound. However, as pointed out in
ref.~\cite{hmty}, there are some regions of parameter space where
partial cancellation between contributions of Feynman diagrams happens. The
cancellation can reduce Br($\tau \to \mu \gamma$) significantly,
e.g., by a factor of $10^{-1}-10^{-2}$, depending on sparticle
mass spectrum. Therefore, in these regions $(\delta^E_{LL})_{23}$
of order one and a low mass spectrum (say, below 1 TeV) are
allowed by the upper bound of Br($\tau \to \mu \gamma$).

\subsection{B Physics}\label{sec3.2}
B meson physics has been able to put good constraint
on $\delta^D_{LL,RR,LR}$ to the
order $10^{-2}$~\cite{ggms,mpr}. These flavor changing structures
contribute mainly through gluino loops. There are also processes,
{\it e.g.}, $b \to s \gamma$, $b \to d(s) l^+ l^-$,
$B \to l^+l^-$ and $B_{d,s}-{\bar B}_{d,s}$ mixing
for which chargino loops contribute a lot. Let's figure
out which entries in $\delta^U$ are relevant to B physics.
First of all, $(\delta^U_{RR})_{ij}$
($i\neq j$) are always irrelevant to the processes because the couplings
involved with the right-handed up-type squarks are proportional to
the Yukawa couplings and their contributions can be neglected
because of the small up and charm Yukawa couplings. Due to the
same reason for $\delta^U_{LR}$, the relevant entries are
$(\delta^U_{LR})_{13,23}$, which refer that the right-handed stop is
propogating in the loop and give us large top Yukawa in the vertex.

Since flipped $SU(5)$ gives predictions on
$(\delta^U_{LR})_{31,32}$ and $\delta^U_{RR}$, not on
$(\delta^U_{LR})_{13,23}$ and $\delta^U_{LL}$, we do not have
strong predictions on the B physics. This is a nice feature of the
model. If flipped $SU(5)$ could give $10^{-2}$ prediction on
$(\delta^U_{LR})_{13}$, the double-penguin contributions to the
$B_d-{\bar B}_d$ mixing would easily be as large as the SM
prediction~\cite{bcrs}. Without these extra flavor mixing entries,
we simply go back to the conclusion made in \cite{bcrs}: the
$\Delta M^{DP}_{B_s}/\Delta M^{SM}_{B_s}$ can be of order 1 while
$\Delta M^{DP}_{B_d}/\Delta M^{SM}_{B_d} \sim \frac{1}{30} \Delta
M^{DP}_{B_s}/\Delta M^{SM}_{B_s}$ and can satisfy the experimental
bounds. Here $\Delta M^{DP}$ means the mass difference given by
the double-penguin diagram and $\Delta M^{SM}$ is the mass
difference given by the SM.

\subsection{$D-{\bar D}$ Mixing}\label{sec3.3}
In the SM, the flavor changing transitions involving external
up-type quarks, which are due to effective flavor changing neutral
current (FCNC) couplings generated at loop level, are much more
suppressed than those involving external down-type quarks. The
effects for external up-type quarks are derived from the virtual
exchanges of down-type quarks in a loop for which GIM
mechanism~\cite{gim} is much more effective because the mass
splittings among down-type quarks are much less than those among
up-type quarks. In the SM model, $D-{\bar D}$ mixing is extremely
small and highly GIM suppressed by the factor $m_s^2/m_W^2 (K_{us}
K_{cs}^*)^2$(of order $10^{-8}$) which makes the SM
prediction of order of $10^6 \hbar s^{-1}$(the contribution from
the bottom loop is smaller because of the smaller CKM mixing).

However, the  GIM mechanism is in general not valid in SUSY
theories. In the scenarios of the minimal flavor violation for
which flavor mixings are described by the CKM mixing, the SUSY
contribution is suppressed by the degeneracy of the first and
second generation squark masses and is of the same order of the SM
contribution. However, with the misalignment of quark and squark
mass matrices in a general SUSY theory, a sizeable
$(\delta^U_{RR,LR})_{12}$ can have significant effects on the
$D-{\bar D}$ mixing through the gluino box diagram. The prediction
on the mass difference, $\Delta M_D$, is proportional to
$(\delta_{12})^2$ for the $CP$ conserving case. The present upper
bound~\cite{pdg} from the CLEO collaboration
\bea 
\Delta M_D < 7
\times 10^{10}~~\textrm{$ \hbar ~s^{-1}$}, ~\textrm{$95\%$ CL}
\label{mdb}
\eea
has been able to put a $10^{-1}-10^{-2}$ constraint on
$(\delta^U_{LL,RR,LR})_{12}$~\cite{cckss} depending on the squark
and gluino mass scale.

In the framework of the constrained MSSM with for example
gravity-mediated SUSY breaking at the high energy scale, we have
indeed the radiatively generated misalignment in the quark and
squark mass matrices as shown in Eq. (\ref{flamssm}). However, the
relevant ${\wti m}^2_{12}$ is too small because it is corrected by
the small entries of the CKM matrix (actually for this entry the
contribution from the strange quark Yukawa coupling is larger but
still too small to be interesting).

In the flipped $SU(5)$ model as we discussed in the subsection \ref{sec2.2},
bilarge neutrino mixing can give $10^{-2}$ right-handed
up-type squark mixing and from which we can get
\bea
\Delta M_D \approx 0.7 \times 10^{9}\times
\frac{(1000 ~\textrm{GeV})^2}{m_{\tilde q}^2}
\times \bigg(\frac{\tan\beta}{50}\bigg)^4 ~\textrm{$\hbar ~s^{-1}$},
\eea
if assuming $m_0 = |A_0| = 2 |M_{1\over 2}|$ which means
$m_{\tilde g}^2/m_{\tilde q}^2 \approx 0.6$. $m_{\tilde q}^2$ in the unit
GeV$^2$ is the averaged right-handed up and charm squark mass squared,
and $m_{\tilde g}$ is the gluino mass. We have used the data
in Eq. (\ref{sol}) as inputs. Typically our prediction is
one or two orders of magnitude smaller than the present bound
which will be accessible at the CLEO-c and BES-III
experiment.

\begin{figure}[t]
\centerline{\psfig{figure=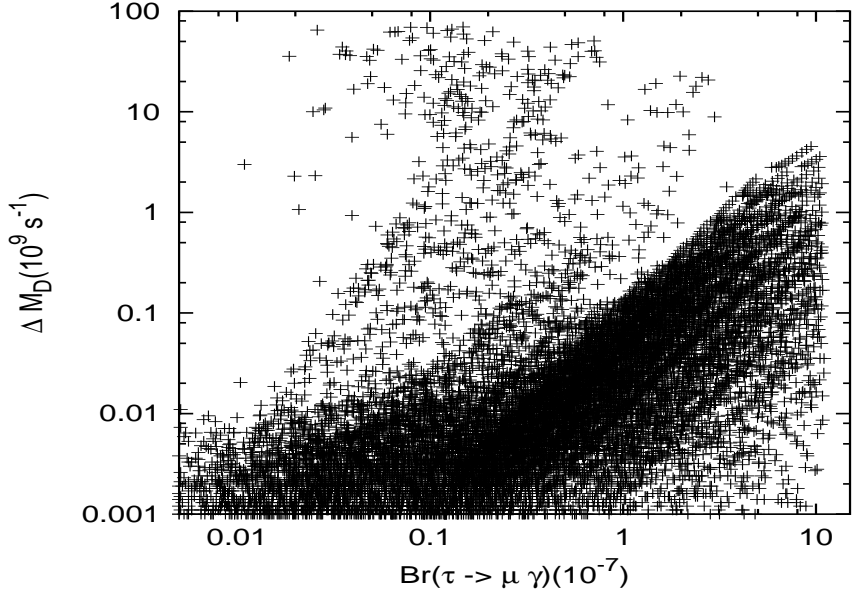,height=8cm,width=12cm}}
\caption{\small The correlation between the $Br(\tau \to \mu
\gamma)$ and $\Delta M_D$ for the parameter space $ 200
~\textrm{GeV} < m_0, |M_{1 \over 2}|, |A_0| < 1.2 ~\textrm{TeV}$,
and $ 10 < \tan\beta <60 $. The constraint $Br(\tau \to \mu
\gamma) < 1.1 \times 10^{-6}$($90 \% ~CL$) has been put.
  }
\label{fig1}
\end{figure}

As shown in the subsection \ref{sec3.1}, the prediction on
$Br(\tau \to \mu \gamma)$ is close to the present bound. Although
$\Delta M_D$ has different dependences on the mass spectrum and
$\tan\beta$ than $Br(\tau \to \mu \gamma)$, we may expect that the
present upper bound on $\Delta M_D$ is hard to be reached in quite
a large part of the parameter space. In the Fig. \ref{fig1}, we
show the correlation between the $Br(\tau \to \mu \gamma)$ and
$\Delta M_D$ in the model. We used the mass insertion approxiamted
formular~\cite{lfv3} to calculate the $\tan\beta$ enhanced
$Br(\tau \to \mu \gamma)$. We used $ 200 ~\textrm{GeV} < m_0,
|M_{1 \over 2}|, |A_0| < 1.2 ~\textrm{TeV}$ and $ 10 < \tan\beta
<60 $. One can see that $\Delta M_D$ can reach $10^9 ~\hbar
s^{-1}$ consistent with the constraint on $Br(\tau \to \mu
\gamma)$. There are also some points for which $\Delta
M_D$ can reach $10^{10} ~\hbar ~s^{-1}$ and $Br(\tau \to \mu
\gamma)$ is smaller than the experimental bound by a factor of
$10^{-1}-10^{-2}$, which corresponds to the case of cancellation
discussed in subsection 3.1. In the plot, we have included the $2
\sigma$ constraints $0.0002<Br(b \to s \gamma)<0.00045$, $\Delta
a^{SUSY}_\mu < 32 \times 10^{-10}$, and the $90\% ~CL$ constraint
$Br(B_s \to \mu^+ \mu^-) < 2.\times 10^{-6}$. Although these
observables are also sensitive to large $\tan\beta$, they are
actually not quite important for the model under study simply
because the strong prediction on $Br(\tau \to \mu \gamma)$ in
quite a large part of the parameter space has constrained the mass
spectrum to be around $1 ~\textrm{TeV}$ and makes other
constraints easily satisfied.

\subsection{Top Physics}\label{sec3.4}

In the SM, the $tc$ transition
rate is also very much GIM suppressed.
The FCNC top quark decays, for example, $t\rightarrow
c V\;\; (V=\gamma, Z, g)$ and $t\rightarrow ch^0$, have branching
ratios 
\bea 
Br(t\rightarrow c \gamma)&\sim& 5\times 10^{-13},\\ 
Br(t\rightarrow c Z)&\sim& 1\times 10^{-13},\\
Br(t\rightarrow c g)&\sim& 4\times 10^{-13},\\ 
Br(t\rightarrow c h^0)&\lsim& 10^{-13}\; ({\rm depended~ on~ m_{h^0}}), 
\eea 
which are
unobservablly small~\cite{smr1,smr2}.

In SUSY theories, GIM mechanism is in general not valid and
sizeable $(\delta^U_{RR,LR})_{23}$ can have significant effects on
the FCNC top to charm transition due to the gluino-mediated
contributions (SUSY-QCD contributions). For $t\rightarrow c V\;\;
(V=\gamma, Z, g)$ and $t\rightarrow ch\;\; (h=h^0, H^0, A^0)$
decays, as pointed out in Ref.~\cite{gs}, by changing
$\delta^U_{23}$ by 3 orders of magnitude the branching ratios
increase by 6 orders of magnitude due to the quadratic dependence
of the branching ratios on the mixing coefficients. Typically for
$\delta^U_{23}\sim 0.4$ and a light sparticle spectrum around 200
GeV, one can get
\bea 
Br(t\rightarrow c h)&\simeq& 10^{-4},\\
Br(t\rightarrow c g)&\lsim& 10^{-5}.
 \eea

In the SM $t \to u$ processes are also GIM suppressed and have
negligible predictions. The radiatively induced $t \to u,c$
couplings by the new physics beyond the SM can be probed by the
top decay processes. Among them $t \to u,c+ h^0$ can be probed by
the top decay. $t \to u, c+ g$ couplings can be probed by the
single top production processes at the Tevatron or LHC: 
\bea 
q {\bar q} \to {\bar u},{\bar c}+t , ~~gg \to {\bar u},{\bar c}+t.
\eea 
There are also direct top quark production processes 
\bea 
u g \to t , ~~ c g \to t. 
\eea 
For the $tug$ coupling, LHC can reach
the sensitivity equivalent to branching ratio $Br(t \to u g) \sim
10^{-6}$. And for the $tcg$ coupling, LHC can reach the
sensitivity equivalent to branching ratio $Br(t \to c g) \sim
10^{-5}$\cite{am}. For $t \to c,u+h^0$, LHC can reach order
$10^{-5}$ branching ratio at $3 \sigma$~\cite{am}.

Our predictions on $\delta^U_{RR,LR}$ are typically of the order
$10^{-2}$ for moderate soft SUSY breaking parameters which is not
large enough to have significant effects and to be observed at the
LHC. In particluar, according to Eq. (\ref{rela}) the present
$D-{\bar D}$ mixing constraint makes $(\delta^U_{RR,LR})_{32,31}$
also constrained in the model under study. However, in some
regions of the parameter space, as pointed out in subsection 3.1,
 we can saturate the $D-{\bar D}$ mixing constraint and
have observable FCNC effects for the top quark physics. In
particular, the $t \to u,c+h^0$ processes are able to reach
$10^{-5}-10^{-6}$ branching ratio which can be observed at LHC for
$100 fb^{-1}$ of integrated luminosity.

 $\delta^U_{RR,LR}$ can give rise to new contributions to more
 processes in top physics. For instance, they will have sizeable
 effects on $t \to c,u+ l^+l^-$ processes.
 These top quark FCNC couplings can also be probed in top and charm
 associated productions at linear and $\gamma-\gamma$
 colliders~\cite{cxy}.

\section{Conclusions and Discussions}
In summary, we have constructed a specific supersymmetric flipped
SU(5) unification model in which bilarge neutrino mixing is
incorporated in the lepton Yukawa couplings. Non-renormalizable
operators have been introduced in the superpotential at the GUT
scale in order to generate the right-handed neutrino Majorana
masses. The effects of other non-renormalizable terms on the low
energy implications can be safely neglected by taking the
theoretical cutoff scale to be the Planck scale. The universal
supersymmetry breaking is assumed at the Planck scale. Because the
up-type and down-type quarks in the model are flipped in the
representations ten and five with respect to the usual SU(5), the
radiatively generated flavor mixing in the squark mass matrix due
to the large neutrino mixing has a pattern different from those in
the conventional SU(5) or SO(10) supersymmetric GUT. This leads to
the phenomenological consequences quite different from SU(5) or
SO(10) supersymmtric GUT. The left-handed slepton mixing is
induced by the up-type quark Yukawa coupling which is not
suppressed when $\tan\beta$ is small. The experimental bound on
$\mu\rightarrow e \gamma$ can be safely satisfied by taking
$\theta_{13}$ small enough. When $\tan\beta$ is large the
branching ratio of the $\tau \to \mu \gamma$ is predicted to reach
the present experimental bound with a mass spectrum around
$1$~TeV. Sizeable radiative corrections happen also to the
right-handed up-type squark mass matrix if $\tan\beta$ is large
($\gsim 30$). This radiatively generated flavor mixing has almost
no impact on B physics. On the contrary, it has effects in top and
charm physics. Because of the special feature of the flipped
$SU(5)$ model, the flavor mixing involving the first generation
up-type squark is never suppressed by small $\theta_{13}$. In the
conventional $SU(5)$ model satisfying the $\mu \to e\gamma$
constraint makes the SUSY prediction on the $K-{\bar K}$ mixing
negligible. However, similar thing never happens in the flipped
$SU(5)$ model. In particular, we have shown that the radiatively
generated falvor mixing can give a promising prediction on the
$D-{\bar D}$ mixing. That is, for moderate values of the soft SUSY
breaking parameters, $\Delta M_D$ is of one order of magnitude
smaller than the present experimental upper bound. Because
$Br(\tau \to \mu \gamma)$ and $\Delta M_D$ have different
dependences on the mass spectrum and $\tan\beta$ they are
basically complementary observables to test the supersymmetric
theory. For the predicted squark flavor mixing, $t \to u,c+h^0$
can reach $10^{-6}-10^{-7}$ branching ratio. In some regions of
the parameter space $\Delta M_D$ can saturate the present bound.
For these ranges of parameter space, $t \to u,c+h^0$ can reach
$10^{-5}-10^{-6}$ which would be observed at the LHC and future
$\gamma-\gamma$ colliders. However a light spectrum is basically
required by having sizeable effects on the top quark FCNC
processes. So the top quark FCNC processes are probably hard to be
found at LHC for the scenario (i) that we have concentrated on in
the paper. A detailed analysis including all the relevant
experimental constraints is needed to make a definite conclusion
on the discovery potential of rare top decays at LHC. Moreover the
model has effects in FCNC D meson decays such as $D\rightarrow
\rho(\pi)\gamma$, which is worth to be examined.

For the scenario (ii) in which only the atmospheric neutrino
mixing is accommodated in the Yukawa coupling, we only have
sizeable predictions on the $(\delta^{U,E}_{RR})_{23,32}$,
 $(\delta^U_{LR})_{32}$ and $(\delta^E_{LR})_{23}$.
 In principle $(\delta^U_{RR,LR})_{32}$ can be quite large (even order one) if
we have very large $A_0$. However, as happened for the scenario
(i) the $\tau \to \mu +\gamma$ would limit them to be smaller than
order of $10^{-1}$ in quite a large part of parameter space and
consequently make the rare top FCNC decay hard to be observed at
the LHC. We need detailed calculations to say definitely on the
discovery potential.

The scenario (iii) is not interesting for 
$D-{\bar D}$ mixing and Top quark FCNC. However it may still
be interesting for the physics of the lepton flavor violation\cite{mvv}.
The point is that by involving a low mass spectrum
the $1/m^4_{\tilde l}$ dependence
of the branching ratio, as shown in Eq. (\ref{approlep}), can compensate 
part of the suppression given by a small off-diagonal
entry in $U_N$. Typically with an order of $10^{-2}$ off-diagonal
entry and a low mass spectrum as for example $m_{\tilde l} \sim 200$ GeV, 
$Br(\tau \to \mu \gamma)$ can still reach $10^{-8}$ and
be accessible in the near future experiments. 

In the paper, we did not consider the possible Majorana phases
in the discussions. These phases, which have no effects in the
neutrino oscillation experiments, can indeed give effects on the
radiative corrections to the squark and slepton mass matrices and
lead to $CP$ violating effects in the low energy phenomena, {\it
e.g.}, the $CP$ violating effects in the lepton physics,
$D-{\bar D}$ mixing and the top quark decay processes. It is
possible that these phases can contribute to the electric dipole
moment of the electron, muon and neutron~\cite{flacp} and are of
much interests.

The flipped $SU(5)$ model can be embedded in the $SO(10)$ model
and we would have similar implications on the flavor physics if
$SO(10)$ takes the breaking chain through the flipped $SU(5)$. We
leave this for the future work.

\section*{Acknowledgements}
L.W would like to thank Vempati Sudhir and Oscar Vives
for helpful communications on the lepton flavor violation.
The research of C.-S. Huang was supported in part by the Natural Science
Foundation of China. And
the research of T. Li was supported  by the National Science Foundation under
 Grant No.~PHY-0070928.
\\
\appendix{\large \bf Appendix A: Some Conventions in MSSM+N}
\newcounter{num}
\setcounter{num}{1}
\setcounter{equation}{0}
\def\theequation{\Alph{num}.\arabic{equation}}

The soft SUSY breaking terms are written as
\bea
-\Delta {\cal L} &=& {\tilde U}^*_i (m^2_U)_{ij} {\tilde U}_j
+{\tilde D}^*_i (m^2_D)_{ij} {\tilde D}_j
+{\tilde Q}^*_i (m^2_Q)_{ij} {\tilde Q}_j
+{\tilde E}^*_i (m^2_E)_{ij} {\tilde E}_j
+{\tilde N}_i^* (m^2_N)_{ij} {\tilde N}_j\nnb \\
&& +{\tilde L}^*_i (m^2_L)_{ij} {\tilde L}_j
+m^2_{H_1} {\tilde H}_1^\dagger {\tilde H}_1
+m^2_{H_2} {\tilde H}_2^\dagger {\tilde H}_2
+\bigg[ {\tilde U}^*_i (Y^A_U)_{ij} {\tilde Q}_j {\tilde H}_2 \nnb \\
&& +{\tilde D}^*_i (Y^A_D)_{ij} {\tilde Q}_j {\tilde H}_1+
{\tilde E}^*_i (Y^A_E)_{ij} {\tilde L}_j {\tilde H}_1
+{\tilde N}_i (Y^A_N)_{ij} {\tilde L}_j {\tilde H}_2
+B \mu {\tilde H}_1 {\tilde H}_2+ h.c. \bigg] ,
\label{softssm1}
\eea
and the gaugino masses are
\bea
-\Delta {\cal L} &=& \frac{1}{2}(M_1 \lambda_1 \lambda_1
+ M_2 \lambda_2 \lambda_2+M_3 \lambda_3 \lambda_3) +h.c.~,
\label{softssm2}
\eea
where ${\tilde Q}$ et.al refer to the scalar parts of the corresponding
superfields. $\lambda_i$($i=1,2,3$) are the gaugino fields for $U(1)_Y$,
$SU(2)_W$ and $SU(3)_C$ groups. In the Appendices B and C, we list the one-loop
RGEs in SUSY model with right-handed neutrinos.

We can parametrize the Yukawa couplings in Eq.(\ref{spssm}) as follows
\bea
Y_U=V_{U_R} y_U V^\dagger_{U_L},~Y_D=V_{D_R} y_D V^\dagger_{D_L},
~Y_E=V_{E_R} y_E V^\dagger_{E_L},~
Y_N=V_{N_R} y_N V^\dagger_{N_L}.
\label{param1}
\eea
$V_{U_R}$ et.al are unitary matrices and $y_U$ et.al are diagonal and
real matrices. After the electroweak symmetry breaking due to the
vacuum expectation values of the $H_1$ and $H_2$ fields
\bea
<H_1> = \pmatrix{0 \cr v_1} ,~~ < H_2 > =\pmatrix{v_2 \cr 0},
\eea
we redefine the
fermion fields to diagonalise the $Y_U$, $Y_D$ and $Y_E$, and
get the Dirac masses of the SM fermions. The CKM matrix is then obtained
in the charged current interactions:
\bea
K=V_{U_L}^\dagger V_{D_L},
\eea
and we define
\bea
U_N=V_{N_L}^{\dagger} V_{E_L}.
\eea
The left-handed neutrino mass matrix in the interaction eigenstate is
\bea
-\Delta{\cal L} &=& \frac{1}{2} {\bar {\nu^c}} M_\nu \nu +h.c. \nnb \\
M_\nu &=& v^2 sin^2\beta V^T_{E_L} Y_N^TM^{-1}_NY_N V_{E_L}
=U^* m_\nu U^\dagger,
\label{neum}
\eea
where $v^2=v_1^2+v_2^2$ and $\tan\beta=v_2/v_1$.
Matrix $U$ is the one responsible for the neutrino mixing.

To see the extra flavor mixing in the squark mass matrix, we redefine the
sfermion fields in the same way
as the corresponding fermion fields and be in the so-called super-CKM base:
\bea
& ({\tilde U_L}, u_L) \to V_{U_L} ({\tilde U_L}, u_L), ~~~
({\tilde U_R}, u_R) \to V_{U_R} ({\tilde U_R}, u_R),~~~ \nnb \\
& ({\tilde D_L}, d_L) \to V_{D_L} ({\tilde D_L}, d_L),~~~
({\tilde D_R}, d_R) \to V_{D_R} ({\tilde D_R}, d_R), ~~~ \nnb \\
& ({\tilde E_L}, e_L) \to V_{E_L} ({\tilde E_L}, e_L), ~~~
({\tilde E_R}, e_R) \to V_{E_R} ({\tilde E_R}, e_R), ~~~
({\tilde \nu_L}, \nu_L) \to V_{E_L} ({\tilde \nu_L}, \nu_L).
\eea
Squark and slepton mass squared matrices will be denoted as ${\wti m}^2$
in the super-CKM base. Trilinear terms will be denoted as
$y^A$ (see Eq. (\ref{softsckm}) for definitions).
The six by six squark mass matrices relevant for the low energy
phenomenology are
\bea
{\wti M}_f^2 = \pmatrix{{\wti M}_{f_{LL}}^2 & {\wti M}_{f_{LR}}^2 \cr
{\wti M}_{f_{LR}}^{2 \dagger} & {\wti M}_{f_{RR}}^2 }, ~~\textrm{ $f=U,D,E$.}
\label{scamass}
\eea
Definitions of the relevant extries can be found in the Appendix B in which
there are additional contributions from the $SU(2)_L\times U(1)_Y$
D-terms
after the electroweak symmetry breaking.

\appendix{\large \bf Appendix B: RGEs in the MSSM+N }
\setcounter{num}{2}
\setcounter{equation}{0}

RGEs are derived as follows
\bea
2 \frac{d {\tilde Y}_U}{dt} &=& (\frac{16}{3} {\tilde \alpha}_3
+3 {\tilde \alpha}_2+\frac{13}{9} \tal_1) \tY_U
-3 (\tY_U \tY_U^\dagger+tr(\tY_U \tY_U^\dagger))\tY_U \nnb \\
&& -\tY_U \tY_D^\dagger \tY_D- tr(\tY_N \tY_N^\dagger) \tY_U, \\
2 \frac{d \tY_D}{dt} &=& (\frac{16}{3}\tal_3+3 \tal_2+\frac{7}{9} \tal_1)\tY_D
-3 (\tY_D^\dagger \tY_D+tr(\tY_D \tY_D^\dagger)) \tY_D\nnb \\
&& - \tY_D \tY_U^\dagger \tY_U- tr(\tY_E \tY_E^\dagger) \tY_D, \\
2 \frac{d \tY_E}{dt} &=&(3 \tal_2+3\tal_1) \tY_E-3 \tY_E \tY_E^\dagger \tY_E
-tr(\tY_E \tY_E^\dagger) \tY_E
-\tY_E \tY_N^\dagger \tY_N -3 tr(\tY_D \tY_D^\dagger) \tY_E,\\
2 \frac{d \tY_N}{dt} &=& (3 \tal_2+\tal_1) \tY_N-3 \tY_N \tY_N^\dagger \tY_N
-tr(\tY_N \tY_N^\dagger)\tY_N
 -\tY_N \tY_E^\dagger \tY_E-3 tr(\tY_U \tY_U^\dagger) \tY_N.
\eea
In the above formulae, we have ${\tilde \alpha}=\alpha/4\pi$,
${\tilde Y}=Y/4\pi$, $t=ln(M^2_{GUT}/Q^2)$.
For the SUSY breaking soft terms, we have
\bea
2 \frac{d \tY_U^A}{dt} &=& (\frac{16}{3} {\tilde \alpha}_3
+3 {\tilde \alpha}_2+\frac{13}{9} \tal_1) \tY_U^A+
2 (\frac{16}{3} {\tilde \alpha}_3 M_3+3 {\tilde \alpha}_2 M_2
+\frac{13}{9} \tal_1 M_1) \tY_U \nnb \\
&& -4 \tY_U \tY_U^\dagger \tY_U^A-6 tr(\tY_U^A \tY_U^\dagger)\tY_U
- 5 \tY_U^A \tY_U^\dagger \tY_U-3 tr(\tY_U \tY_U^\dagger) \tY_U^A \nnb \\
&& -2 \tY_U \tY_D^\dagger \tY_D^A -\tY_U^A \tY_D^\dagger \tY_D
-2 tr(\tY_N^A \tY_N^\dagger) \tY_U-tr(\tY_N \tY_N^\dagger) \tY_U^A,\\
2 \frac{ d \tY_D^A}{dt} &=& (\frac{16}{3} {\tilde \alpha}_3
+3 {\tilde \alpha}_2+\frac{7}{9} \tal_1) \tY_D^A+
2 (\frac{16}{3} {\tilde \alpha}_3 M_3+3 {\tilde \alpha}_2 M_2
+\frac{7}{9} \tal_1 M_1) \tY_D \nnb \\
&& -4 \tY_D \tY_D^\dagger \tY_D^A-6 tr(\tY_D^A \tY_D^\dagger)\tY_D
- 5 \tY_D^A \tY_D^\dagger \tY_D-3 tr(\tY_D \tY_D^\dagger) \tY_D^A \nnb \\
&& -2 \tY_D \tY_U^\dagger \tY_U^A -\tY_D^A \tY_U^\dagger \tY_U
-2 tr(\tY_E^A \tY_E^\dagger) \tY_D-tr(\tY_E \tY_E^\dagger) \tY_D^A,\\
2 \frac{d \tY_E^A}{dt} &=& (3 \tal_2+3\tal_1) \tY_E^A+
2 (3 \tal_2 M_2 +3\tal_1 M_1) \tY_E \nnb \\
&& -4 \tY_E \tY_E^\dagger \tY_E^A-2 tr(\tY_E^A \tY_E^\dagger) \tY_E
-5 \tY_E^A \tY_E^\dagger \tY_E- tr(\tY_E \tY_E^\dagger) \tY_E^A \nnb \\
&& -2 \tY_E \tY_N^\dagger \tY_N^A-\tY_E^A \tY_N^\dagger \tY_N
-6 tr(\tY_D^A \tY_D^\dagger) \tY_E -3 tr(\tY_D \tY_D^\dagger) \tY_E^A,\\
2 \frac{d \tY_N^A}{dt} &=& (3 \tal_2+\tal_1) \tY_N^A
+2 ( 3\tal_2 M_2 +\tal_1 M_1) \tY_N \nnb \\
&& -4 \tY_N \tY_N^\dagger \tY_N^A- 2 tr(\tY_N^A \tY_N^\dagger) \tY_N
-5 \tY_N^A \tY_N^\dagger \tY_N- tr(\tY_N \tY_N^\dagger) \tY_N^A \nnb \\
&& -2 \tY_N \tY_E^\dagger \tY_E^A-\tY_N^A \tY_E^\dagger \tY_E
-6 tr(\tY_U^A \tY_U^\dagger) \tY_N -3 tr(\tY_U \tY_U^\dagger) \tY_N^A,\\
\frac{d m^2_U}{dt} &=&(\frac{16}{3} \tal_3 |M_3|^2+\frac{16}{9} \tal_1 |M_1|^2)
-(\tY_U \tY_U^\dagger m_U^2+m_U^2 \tY_U \tY_U^\dagger) \nnb \\
&& -2 ( \tY_U m_Q^2 \tY_U^\dagger +\tY_U \tY_U^\dagger m_{H_2}^2
+\tY_U^A \tY_U^{A \dagger}), \\
\frac{d m^2_D}{dt} &=&(\frac{16}{3} \tal_3 |M_3|^2+\frac{4}{9} \tal_1 |M_1|^2)
-(\tY_D \tY_D^\dagger m_D^2+m_D^2 \tY_D \tY_D^\dagger) \nnb \\
&& -2 ( \tY_D m_Q^2 \tY_D^\dagger +\tY_D \tY_D^\dagger m_{H_1}^2
+\tY_D^A \tY_D^{A \dagger}), \\
\frac{d m^2_Q}{dt} &=& (\frac{16}{3} \tal_3 |M_3|^2+ 3 \tal_2 |M_2|^2
+ \frac{1}{9} \tal_1 |M_1|^2) \nnb \\
&& -\frac{1}{2}(\tY_U^\dagger \tY_U m_Q^2+m_Q^2 \tY_U^\dagger \tY_U)
-( \tY_U^\dagger m_U^2 \tY_U+\tY_U^\dagger \tY_U m_{H_2}^2
+ \tY_U^{A\dagger} \tY_U^A) \nnb \\
&& -\frac{1}{2}(\tY_D^\dagger \tY_D m_Q^2+m_Q^2 \tY_D^\dagger \tY_D)
-( \tY_D^\dagger m_D^2 \tY_D+\tY_D^\dagger \tY_D m_{H_1}^2
+ \tY_D^{A\dagger} \tY_D^A) , \\
\frac{d m_N^2}{dt} &=& -\bigg[m_N^2 (\tY_N \tY_N^\dagger)^T
+(\tY_N \tY_N^\dagger)^T m_N^2 \bigg]
-2 (\tY_N m_L^2 \tY_N^\dagger+\tY_N \tY_N^\dagger m_{H_2}^2
+\tY_N^A \tY_N^{A \dagger})^T, \\
\frac{d m_E^2}{dt} &=& 4 \tal_1 |M_1|^2
-(m_E^2 \tY_E \tY_E^\dagger +\tY_E \tY_E^\dagger m_E^2)
 -2 (\tY_E m_L^2 \tY_E^\dagger+\tY_E \tY_E^\dagger m_{H_1}^2
+\tY_E^A \tY_E^{A \dagger}), \\
\frac{d m_L^2}{dt} &=& (3 \tal_2 |M_2|^2+\tal_1 |M_1|^2) \nnb \\
&& -\frac{1}{2}(\tY_N^\dagger \tY_N m_L^2+m_L^2 \tY_N^\dagger \tY_N)
-( \tY_N^\dagger (m_N^2)^T \tY_N+\tY_N^\dagger \tY_N m_{H_2}^2
+ \tY_U^{A\dagger} \tY_U^A) \nnb \\
&& -\frac{1}{2}(\tY_E^\dagger \tY_E m_L^2+m_L^2 \tY_E^\dagger \tY_E)
-( \tY_E^\dagger m_E^2 \tY_E+\tY_E^\dagger \tY_E m_{H_1}^2
+ \tY_E^{A\dagger} \tY_E^A) , \\
\frac{d m_{H_2}^2}{dt} &=& (3 \tal_2 |M_2|^2+\tal_1 |M_1|^2) \nnb \\
&& -3 tr( \tY_U^\dagger \tY_U m_{H_2}^2+\tY_U^\dagger m_U^2 \tY_U
+\tY_U m_Q^2 \tY_U^\dagger +\tY_U^A \tY_U^{A \dagger}) \nnb \\
&& - tr( \tY_N^\dagger \tY_N m_{H_2}^2+\tY_N^\dagger (m_N^2)^T \tY_N
+\tY_N m_L^2 \tY_N^\dagger +\tY_N^A \tY_N^{A \dagger}), \\
\frac{d m_{H_1}^2}{dt} &=& (3 \tal_2 |M_2|^2+\tal_1 |M_1|^2) \nnb \\
&& -3 tr( \tY_D^\dagger \tY_D m_{H_1}^2+\tY_D^\dagger m_D^2 \tY_D
+\tY_D m_Q^2 \tY_D^\dagger +\tY_D^A \tY_D^{A \dagger}) \nnb \\
&& - tr( \tY_E^\dagger \tY_E m_{H_1}^2+\tY_E^\dagger m_E^2 \tY_E
+\tY_E m_L^2 \tY_E^\dagger +\tY_E^A \tY_E^{A \dagger}).
\eea

In the super-CKM base, the Yukawa couplings of the SM fermions are all
diagonalised and generate the SM fermions in the mass eigenstates.
The SUSY breaking soft terms in this base can be
written as
\bea
&& {\wti m}^2_{U_L}= V^\dagger_{U_L} m^2_Q V_{U_L}, ~~
{\wti m}^2_{D_L}= V^\dagger_{D_L} m^2_Q V_{D_L}, ~~
{\wti m}^2_{U_R}= V^\dagger_{U_R} m^2_U V_{U_R}, ~~ \nnb \\
&& {\wti m}^2_{D_R}= V^\dagger_{D_R} m^2_D V_{D_R}, ~~
{\wti m}^2_{E_L}= {\wti m}^2_{\nu_L}= V^\dagger_{E_L} m^2_L V_{E_L}, ~~
{\wti m}^2_{E_R}= V^\dagger_{E_R} m^2_E V_{E_R},~~ \nnb \\
&& y^A_U=V^\dagger_{U_R} Y^A_U V_{U_L}, ~~
y^A_D=V^\dagger_{D_R} Y^A_D V_{D_L}, ~~
y^A_E=V^\dagger_{E_R} Y^A_E V_{E_L}.
\label{softsckm}
\eea
Trilinear terms are transformed similar to the Yukawa couplings.
The left-handed neutrinos in this base are called as interaction
eigenstate neutrinos. Together with the D-term contributions after the
electroweak symmetry breaking, ${\wti m}^2$ and $y^A$ give
the mass matrices of sfermions, Eq. (\ref{scamass}), relevant
to the low energy phenomenology. The entries in Eq. (\ref{scamass}) are
\bea
{\wti M}_{U_{LL}}^2&=&{\wti m}_{U_L}^2+ m^2_U+ M_Z^2 cos2\beta(\frac{1}{2}-\frac{2}{3}
sin^2\theta_W), ~~ \nnb \\
{\wti M}_{U_{RR}}^2 &=& {\wti m}_{U_R}^2+m^2_U+M_Z^2 cos2\beta(\frac{2}{3}
sin^2\theta_W), ~~ \nnb \\
{\wti M}_{U_{LR}}^2 &=& -v_2 (y_U^{A \dagger}+ \mu \cot\beta y_U) ; \\
{\wti M}_{D_{LL}}^2 &=&{\wti m}_{D_L}^2+m^2_D- M_Z^2 cos2\beta(\frac{1}{2}-\frac{1}{3}
sin^2\theta_W), ~~ \nnb \\
{\wti M}_{D_{RR}}^2 &=&{\wti m}_{D_R}^2+m^2_D-M_Z^2 cos2\beta(\frac{1}{3}
sin^2\theta_W), ~~ \nnb \\
{\wti M}_{D_{LR}}^2 &=& -v_1 (y_D^{A \dagger}+ \mu \tan\beta y_D);\\
{\wti M}_{E_{LL}}^2 &=&{\wti m}_{E_L}^2+m^2_E- M_Z^2 cos2\beta(\frac{1}{2}-
sin^2\theta_W), ~~ \nnb \\
{\wti M}_{E_{RR}}^2 &=&{\wti m}_{E_R}^2+m^2_E-M_Z^2 cos2\beta
sin^2\theta_W, ~~ \nnb \\
{\wti M}_{E_{LR}}^2 &=& -v_1 (y_E^{A \dagger}+ \mu \tan\beta y_E).
\eea
$m_U$, $m_D$ and $m_E$ are the diagonal SM fermion mass matrices.

Assuming the SUSY breaking is universal at $M_{GUT}$ scale, we
are able to get the radiatively corrected SUSY breaking soft terms.
At the first order, the radiative corrections given by the Yukawa
and trilinear couplings are
\bea
\delta_1 {\wti m}_{U_R}^2 &=& -\frac{2}{16\pi^2} (3 m_0^2+|A_0|^2) y_U^2
\Delta_1 t, \\
\delta_1 {\wti m}_{D_R}^2 &=& -\frac{1}{16\pi^2} (3 m_0^2+|A_0|^2) y_D^2
\Delta_1 t, \\
\delta_1 {\wti m}_{U_L}^2 &=& -\frac{1}{16\pi^2} (3 m_0^2+|A_0|^2)
( y_U^2 + K y_D^2 K^\dagger) \Delta_1 t,\\
\delta_1 {\wti m}_{D_L}^2 &=& -\frac{1}{16\pi^2} (3 m_0^2+|A_0|^2)
( K^\dagger y_U^2 K+ y_D^2 ) \Delta_1 t,\\
\delta_1 {\wti m}_{E_R}^2 &=& -\frac{2}{16 \pi^2} (3 m_0^2+|A_0|^2)
y_E^2 \Delta_1 t,\\
\delta_1 {\wti m}_{E_L}^2 &=& -\frac{1}{16\pi^2} (3 m_0^2+|A_0|^2)
( y_E^2 \Delta_1 t+U_N^\dagger y_N^2 U_N
{\bar \Delta}_1 t)=\delta_1 {\wti m}_{\nu_L}^2, \\
\delta_1 y_U^A &=&-\frac{3}{32 \pi^2} A_0 \bigg[ [3 y_U( y_U^2+tr(y_U^2))
+y_U K y_D^2 K^\dagger ] \Delta_1 t
+ y_U tr(y_N^2) {\bar \Delta}_1 t \bigg],\\
\delta_1 y_D^A &=& -\frac{3}{32 \pi^2} A_0 \bigg[ 3 y_D( y_D^2+tr(y_D^2))
+y_D K^\dagger y_U^2 K + y_D tr(y_E^2)] \Delta_1 t,\\
\delta_1 y_E^A &=& -\frac{3}{32 \pi^2} A_0 \bigg[ [y_E^3+ y_E tr(y_E^2)
+3 y_E tr(y_D^2) ] \Delta_1 t
+ y_E U_N^\dagger y_N^2 U_N {\bar \Delta}_1 t \bigg],\\
\Delta_1 t &=& ln\frac{M_{GUT}^2}{M_{SUSY}^2}, ~~~
{\bar \Delta}_1 t =ln\frac{M_{GUT}^2}{M_N^2},
\eea
where we assume the right-handed neutrino fields $N_i$ decouple at the
same scale. The above estimates are presented at the first order. A detailed
analysis beyond the first order has been presented in~\cite{ach} for
MSSM. The results are in agreement with the qualitative expectations
based on the first order approximation. In the model with
the right-handed neutrino fields, similar conclusion can be obtained.
\\

\appendix{\large \bf Appendix C: RGEs in the Flipped $SU(5)$ Model} \nonumber
\setcounter{num}{3}
\setcounter{equation}{0}

In the flipped $SU(5)$ model, the RGEs are obtained as follows
\bea
\frac{d {\tilde Y}_{10}}{dt}&=&(2 C_{10}+C_h){\tilde \alpha}_5 {\tilde Y}_{10}
+(2 C^X_{10}+C^X_h) {\tilde \alpha}_X {\tilde Y}_{10}
- 3 {\tilde Y}_{10}{\tilde Y}_{10}^\dagger {\tilde Y}_{10} \nnb \\
&& -\bigg[{\tilde Y}_{\bar 5}{\tilde Y}_{\bar 5}^\dagger {\tilde Y}_{10}+
{\tilde Y}_{10} ({\tilde Y}_{\bar 5}{\tilde Y}_{\bar 5}^\dagger)^T \bigg]
-\frac{1}{2} \bigg[3 tr({\tilde Y}_{10}{\tilde Y}_{10}^\dagger)
+tr({\tilde Y}_1{\tilde Y}_1^\dagger) \bigg] {\tilde Y}_{10}, \\
\frac{d {\tilde Y}_{\bar 5}}{dt}&=& (C_{10}+C_5+C_h){\tilde \alpha}_5
{\tilde Y}_{\bar 5}+(C^X_{10}+C^X_5+C^X_h) {\tilde \alpha}_X {\tilde Y}_{\bar 5}
\nnb \\
&&-\frac{3}{2}{\tilde Y}_{10} {\tilde Y}_{10}^\dagger {\tilde Y}_{\bar 5}
-3 {\tilde Y}_{\bar 5}{\tilde Y}_{\bar 5}^\dagger {\tilde Y}_{\bar 5}
-\frac{1}{2}{\tilde Y}_{\bar 5}{\tilde Y}_1^\dagger {\tilde Y}_1
-2 tr({\tilde Y}_{\bar 5}{\tilde Y}_{\bar 5}^\dagger) {\tilde Y}_{\bar 5},\\
\frac{d {\tilde Y}_1}{dt}&=& (C_5+C_h) {\tilde \alpha}_5{\tilde Y}_1
+(C^X_5+C^X_h+C^X_1) {\tilde \alpha}_X {\tilde Y}_1 \nnb \\
&& -3{\tilde Y}_1{\tilde Y}_1^\dagger {\tilde Y}_1
 -2{\tilde Y}_1{\tilde Y}_{\bar 5}^\dagger {\tilde Y}_{\bar 5}
-\frac{1}{2}\bigg[ tr({\tilde Y}_1{\tilde Y}_1^\dagger)
+3 tr({\tilde Y}_{10}{\tilde Y}_{10}) \bigg] {\tilde Y}_1.
\eea
Here we have $t=ln\frac{M^2_*}{Q^2}$ and the following
Casimir operators:
\bea
C_{10}=\frac{18}{5}, ~~C_h=C_5=\frac{12}{5},~~C^X_{10}=\frac{1}{40},~~
C^X_5=\frac{9}{40},~~C^X_1=\frac{5}{8},~~C^X_h=\frac{1}{10}.
\label{casimirop}
\eea
$\alpha_5$ and $\alpha_X$ are the gauge couplings of
$SU(5)$ and $U(1)_X$ gauge groups, respectively.
For the SUSY breaking soft terms,
we have
\bea
\frac{d {\tilde Y}^A_{10}}{dt}&=&(2 C_{10}+C_h) {\tilde \alpha}_5
({\tilde Y}^A_{10} +2  M_5 {\tilde Y}_{10})
+(2 C^X_{10}+C^X_h){\tilde \alpha}_X ({\tilde Y}^A_{10} +2 M_X {\tilde Y}_{10})
\nnb \\
&& -\frac{9}{2}({\tilde Y}_{10}^A{\tilde Y}_{10}^\dagger {\tilde Y}_{10}
+{\tilde Y}_{10}{\tilde Y}_{10}^\dagger {\tilde Y}_{10}^A)
-2\bigg[{\tilde Y}_{\bar 5}^A{\tilde Y}_{\bar 5}^\dagger {\tilde Y}_{10}
+{\tilde Y}_{10}({\tilde Y}_{\bar 5}^A{\tilde Y}_{\bar 5}^\dagger)^T \bigg]
-\bigg[{\tilde Y}_{\bar 5}{\tilde Y}_{\bar 5}^\dagger {\tilde Y}_{10}^A
+{\tilde Y}_{10}^A({\tilde Y}_{\bar 5}{\tilde Y}_{\bar 5}^\dagger)^T \bigg]
\nnb \\
&& -\frac{3}{2}\bigg[2 tr({\tilde Y}_{10}^A{\tilde Y}_{10}^\dagger)
{\tilde Y}_{10}
+tr({\tilde Y}_{10}{\tilde Y}_{10}^\dagger){\tilde Y}_{10}^A \bigg]
-\frac{1}{2} \bigg[ 2 tr({\tilde Y}_1^A{\tilde Y}_1^\dagger){\tilde Y}_{10}
+tr({\tilde Y}_1{\tilde Y}_1^\dagger){\tilde Y}_{10}^A \bigg], \\
\frac{d {\tilde Y}_{\bar 5}^A}{dt} &=&(C_{10}+C_5+C_h){\tilde \alpha}_5
({\tilde Y}_{\bar 5}^A+2 M_5 {\tilde Y}_{\bar 5})+ (C_{10}^X+C_5^X+C_h^X)
{\tilde \alpha}_X( {\tilde Y}_{\bar 5}^A +2 M_X {\tilde Y}_{\bar 5}) \nnb \\
&& -\frac{3}{2}{\tilde Y}_{10} {\tilde Y}_{10}^\dagger {\tilde Y}_{\bar 5}^A
-3 {\tilde Y}_{10}^A {\tilde Y}_{10}^\dagger {\tilde Y}_{\bar 5}
-4 {\tilde Y}_{\bar 5}^A{\tilde Y}_{\bar 5}^\dagger {\tilde Y}_{\bar 5}
-5 {\tilde Y}_{\bar 5}{\tilde Y}_{\bar 5}^\dagger {\tilde Y}_{\bar 5}^A \nnb \\
&& -\frac{1}{2} {\tilde Y}_{\bar 5}^A {\tilde Y}_1^\dagger {\tilde Y}_1
- {\tilde Y}_{\bar 5} {\tilde Y}_1^\dagger {\tilde Y}_1^A
-2 tr({\tilde Y}_{\bar 5} {\tilde Y}_{\bar 5}^\dagger) {\tilde Y}_{\bar 5}^A
-4 tr({\tilde Y}_{\bar 5}^A {\tilde Y}_{\bar 5}^\dagger) {\tilde Y}_{\bar 5} \\
\frac{d {\tilde Y}_1^A}{dt}& =& (C_5+C_h) {\tilde \alpha}_5 ({\tilde Y}_1^A
+2 M_5 {\tilde Y}_1)+(C_5^X+C^X_h+C^X_1) {\tilde \alpha}_X
({\tilde Y}_1^A+2 M_X {\tilde Y}_1)  \nnb \\
&& -\frac{7}{2} {\tilde Y}_1 {\tilde Y}_1^\dagger {\tilde Y}_1^A
-\frac{11}{2} {\tilde Y}_1^A {\tilde Y}_1^\dagger {\tilde Y}_1
-2 {\tilde Y}_1^A {\tilde Y}_{\bar 5}^\dagger {\tilde Y}_{\bar 5}
-4 {\tilde Y}_1 {\tilde Y}_{\bar 5}^\dagger {\tilde Y}_{\bar 5}^A \nnb \\
&& -\frac{1}{2}\bigg[ tr({\tilde Y}_1 {\tilde Y}_1^\dagger) {\tilde Y}_1^A
+2 tr({\tilde Y}_1^A {\tilde Y}_1^\dagger) {\tilde Y}_1 \bigg]
-\frac{3}{2} \bigg[ tr({\tilde Y}_{10}{\tilde Y}_{10}^\dagger) {\tilde Y}_1^A
+2 tr({\tilde Y}_{10}^A {\tilde Y}_{10}^\dagger) {\tilde Y}_1\bigg], \\
\frac{d m^2_{10}}{dt} &=& 4 C_{10} {\tilde \alpha}_5 |M_5|^2
+4 C^X_{10} {\tilde \alpha}_X |M_X|^2 -\frac{3}{2}\bigg[ m^2_{10}
({\tilde Y}_{10} {\tilde Y}_{10}^\dagger)^T
+{\tilde Y}_{10} {\tilde Y}_{10}^\dagger m^2_{10} \bigg] \nnb \\
&& -\bigg[ m^2_{10} ({\tilde Y}_{\bar 5} {\tilde Y}_{\bar 5}^\dagger)^T
+({\tilde Y}_{\bar 5} {\tilde Y}_{\bar 5}^\dagger)^T m^2_{10} \bigg]
-3 ( {\tilde Y}_{10}^A {\tilde Y}_{10}^{A \dagger}
+{\tilde Y}_{10} m^2_{10} {\tilde Y}_{10}^\dagger)^T \nnb \\
&& -2 ({\tilde Y}_{\bar 5}^A {\tilde Y}_{\bar 5}^{A \dagger}+
{\tilde Y}_{\bar 5} m^2_{\bar 5} {\tilde Y}_{\bar 5}^\dagger)^T
-3 ({\tilde Y}_{10} {\tilde Y}_{10}^\dagger)^T m^2_h
-2 ({\tilde Y}_{\bar 5} {\tilde Y}_{\bar 5}^\dagger)^T m^2_{\bar h} \\
\frac{d m^2_{\bar 5}}{dt} &=& 4 C_5 {\tilde \alpha}_5|M_5|^2
+4 C^X_5 {\tilde \alpha}_X |M_X|^2
-2 ( m^2_{\bar 5} {\tilde Y}_{\bar 5}^\dagger {\tilde Y}_{\bar 5}
+{\tilde Y}_{\bar 5}^\dagger {\tilde Y}_{\bar 5} m^2_{\bar 5})
-\frac{1}{2} (m^2_{\bar 5} {\tilde Y}_1^\dagger {\tilde Y}_1+
{\tilde Y}_1^\dagger {\tilde Y}_1 m^2_{\bar 5}) \nnb \\
&& -4 ( {\tilde Y}_{\bar 5}^\dagger (m^2_{10})^T {\tilde Y}_{\bar 5}
+{\tilde Y}_{\bar 5}^{A \dagger} {\tilde Y}_{\bar 5}^A)
-({\tilde Y}_1^\dagger (m^2_1)^T {\tilde Y}_1
+{\tilde Y}_1^{A \dagger} {\tilde Y}_1^A)
-4 {\tilde Y}_{\bar 5}^\dagger {\tilde Y}_{\bar 5} m^2_{\bar h}
-{\tilde Y}_1^\dagger {\tilde Y}_1 m^2_h, \\
\frac{d m^2_1}{dt} &=& 4 {\tilde \alpha}_X C_1^X |M_X|^2
-\frac{5}{2}( ({\tilde Y}_1 {\tilde Y}_1^\dagger)^T m^2_1+
m^2_1 ({\tilde Y}_1 {\tilde Y}_1^\dagger)^T) \nnb \\
&& -5 ({\tilde Y}_1 m^2_{\bar 5} {\tilde Y}_1^\dagger+
{\tilde Y}_1^A {\tilde Y}_1^{A \dagger})^T
-5 ({\tilde Y}_1 {\tilde Y}_1^\dagger)^T m^2_h, \\
\frac{d m^2_h}{dt} &=& 4 C_h {\tilde \alpha}_5 |M_5|^2
+4 C^X_h {\tilde \alpha}_X |M_X|^2 -3 tr[{\tilde Y}_{10} {\tilde Y}_{10}^\dagger
m^2_h +2 {\tilde Y}_{10} m^2_{10} {\tilde Y}_{10}^\dagger
+{\tilde Y}_{10}^A {\tilde Y}_{10}^{A \dagger}] \nnb \\
&& - tr[ {\tilde Y}_1 {\tilde Y}_1^\dagger m^2_h
+ {\tilde Y}_1 m^2_{\bar 5} {\tilde Y}_1^\dagger
+{\tilde Y}_1^T m^2_1 {\tilde Y}_1^*+{\tilde Y}_1^A {\tilde Y}_1^{A \dagger}]\\
\frac{d m^2_{\bar h}}{dt} &=& 4 C_h {\tilde \alpha}_5 |M_5|^2
+4 C^X_h{\tilde \alpha}_X |M_X|^2 -4 tr[{\tilde Y}_{\bar 5}
{\tilde Y}_{\bar 5} ^\dagger m^2_{\bar h}+{\tilde Y}_{\bar 5} m^2_{\bar 5}
{\tilde Y}_{\bar 5}^\dagger +{\tilde Y}_{\bar 5}^T m^2_{10}
{\tilde Y}_{\bar 5}^*+{\tilde Y}_{\bar 5}^A {\tilde Y}_{\bar 5}^{A \dagger}].
\eea

According to the above RGEs, we can estimate the first order
radiative corrections to the SUSY breaking soft terms. Using
Eqs. (\ref{softsckm}),
(\ref{bound1}) and (\ref{bound2}), we can write them in the super-CKM
base and see clearly the extra flavor structure. Assuming the universal
SUSY breaking soft terms, $m_0$ for scalars, $M_{1 \over 2}$ for gauginos
and $Y^A = A_0 Y$ at the $M_*$ scale, we get the following corrections
given by the Yukawa and trilinear couplings
\bea
\delta_2 {\wti m}^2_{D_L} &=&-\frac{1}{16 \pi^2}\bigg[3(3 m^2_0+|A_0|^2) y_D^2
+2 (3 m^2_0+|A_0|^2) K^\dagger y_U^2 K \bigg] \Delta_2 t, \\
\delta_2 {\wti m}^2_{U_L} &=& -\frac{1}{16 \pi^2} \bigg[3(3 m^2_0+|A_0|^2)
K y_D^2 K^\dagger+2 (3 m^2_0+|A_0|^2)y_U^2 \bigg] \Delta_2 t, \\
\delta_2 {\wti m}^2_{U_R} &=&-\frac{1}{16 \pi^2}\bigg[4(3m^2_0+|A_0|^2) y_U^2
+(3 m^2_0+|A_0|^2) U^*_N y_E^2 U^T_N \bigg] \Delta_2 t,\\
\delta_2 {\wti m}^2_{D_R} &=&-\frac{1}{16 \pi^2}\bigg[3(3 m^2_0+|A_0|^2)y_D^2
+2 (3 m^2_0+|A_0|^2) K^T y_U^2 K^*) \bigg] \Delta_2 t,\\
\delta_2 {\wti m}^2_{E_L}&=&-\frac{1}{16 \pi^2} \bigg[(3 m^2_0+|A_0|^2) y_E^2
+ 4 (3 m^2_0+|A_0|^2) U^\dagger_N y_U^2 U_N \bigg]\Delta_2 t,\\
\delta_2 {\wti m}^2_{E_R} &=& -\frac{1}{16 \pi^2} 5(3 m^2_0+|A_0|^2) y_E^2
\Delta_2 t, ~~~\delta m^2_{\nu_L}=\delta m^2_{E_L}, \\
\delta_2 y^A_U &=& -\frac{A_0}{16 \pi^2}\bigg[\frac{9}{2} y_U K y_D^2 K^\dagger
+9 y_U^3 +\frac{3}{2} U_N^* y_E^2 U_N^T y_U+6 tr(y_U^2) y_U \bigg]\Delta_2 t,\\
\delta_2 y^A_D &=& -\frac{A_0}{16 \pi^2}\bigg[9 y_D^3 +3 (y_D K^\dagger y_U^2 K
+K^T y_U^2 K^* y_D) +\frac{3}{2}(tr(y_E^2)+3 tr(y_D^2)) y_D \bigg] \Delta_2 t,\\
\delta_2 y^A_E &=& -\frac{A_0}{16 \pi^2}\bigg[ 9 y_E^3+ 5 y_E U_N^\dagger
y_U^2 U_N+\frac{3}{2}(tr(y_E^2)+3 tr(y_D^2)) y_E \bigg] \Delta_2 t,\\
&& \Delta_2 t=ln\frac{M^2_*}{M^2_{GUT}}.
\eea


\begin{thebibliography}{99}
\bibitem{skatm}
Y. Fukuda et al. [Super-Kamiokande Collaboration],
Phys. Rev. Lett. {\bf 81}(1998)1562.
\bibitem{sno}
Q.R. Ahmad et al.  [SNO Collaboration],
Phys. Rev. Lett. {\bf 89}(2002)011301;
Phys. Rev. Lett. {\bf 89}(2002)011302.
\bibitem{kamland}
K. Eguchi et al. [KamLAND Collaboration],
Phys. Rev. Lett. {\bf 90}(2003)021802.
\bibitem{chooz} M. Apollonio et al.[CHOOZ Collaboration],
Phys.Lett.B{\bf 466}(1999)415.
\bibitem{shiozawa}M. Shiozawa, talk given at the ICHEP2002 for the
Super-K collaboration;
\bibitem{phs} P.C. de Holanda and A.Yu. Smirnov, hep-ph/0212270.
\bibitem{also} see also M. Maltoni, T. Schwetz, M.A. Tortola,
J.W.F. Valle,  Phys. Rev. D{\bf 67}(2003)013011[hep-ph/0207227];
S. Pakvasa, J. W. F. Valle, hep-ph/0301061.
\bibitem{ponte} B. Pontecorva, Sov. Phys. JETP {\bf 6}(1958)429.
\bibitem{Maki:1962mu}
Z. Maki, M. Nakagawa and S. Sakata, Prog. Theor. Phys. {\bf 28}, 870 (1962).
\bibitem{msw} L. Wolfenstein, Phys. Rev. D{\bf 17}(1978)2369;
S.P. Mikheev and A.Yu. Smirnov, Sov. J. Nucl. Phys. {\bf 42}(1985)913
, Yad. Fiz. {\bf 42}(1985)1441.
\bibitem{ss}
T.~Yanagida, in {\it Proc. of the Workshop on the
    Unified Theory and Baryon Number in the Universe}, ed. O.~Sawada
  and A.~Sugamoto
  (KEK report 79-18, 1979), p. 95; \\
  M.~Gell-Mann, P.~Ramond and R.~Slansky, in {\it Supergravity}, ed.
  P.~van Nieuwenhuizen and D.Z.~Freedman (North Holland, Amsterdam,
  1979), p. 315; E.~Witten,
%``Neutrino Masses In The Minimal O(10) Theory,''
Phys.\ Lett.\ B {\bf 91}, 81 (1980);
R.~N.~Mohapatra and G.~Senjanovic,
%``Neutrino Mass And Spontaneous Parity Nonconservation,''
Phys.\ Rev.\ Lett.\  {\bf 44}, 912 (1980);\\
T.~Yanagida,
%``Horizontal Gauge Symmetry And Masses Of Neutrinos,''
Prog.\ Theor.\ Phys.\  {\bf 64}, 1103 (1980).
\bibitem{wwzf} S. Weinberg, Trans.N.Y.Acad.Sci.{\bf 38}(1977)185;
F. Wilczek and A. Zee, Phys. Lett. B{\bf 70}(1977)418;
H. Fritzsch, Phys. Lett. B{\bf 70}(1977)436.
\bibitem{FN}
C. D. Froggatt and H. B. Nielsen, Nucl. Phys. {\bf B}147 (1979) 277.
\bibitem{asy} K.S. Babu and S.M. Barr, Phys. Lett. B {\bf 381}(1996) 202;
         C.H. Albright, K.S. Babu, and S.M. Barr, Phys. Rev. Lett.
         {\bf 81}(1998)1167; J. Sato and T. Yanagida,
         Phys. Lett. B {\bf 430}(1998)127; N. Irges, S. Lavignac,
         and P. Ramond, Phys. Rev. D {\bf 58}1998)035003.
\bibitem{bdv}W. Buchmuller, D. ~Delepine and F. ~Vissani, Phys. Lett.
{\bf B459} 171 (1999); W. Buchmuller, D. Delepine and L. T.
Handoko, Nucl. Phys.\ {\bf B576} 445 (2000); J. Ellis, M. E.
Gomez, G. K. Leontaris, S. Lola and D. V. Nanopoulos, Eur. Phys.
J. {\bf C14} 319 (2000); J. Hisano and K. Tobe, Phys. Lett. {\bf
B510} 197 (2001); J. A. Casas and A. Ibarra, arXiv:hep-ph/0103065;
D. F. Carvalho, J. Ellis, M. E. Gomez and S. Lola,
arXiv:hep-ph/0103256; T. Blazek and S. F. King,
arXiv:hep-ph/0105005; J. Sato, K. Tobe and T. Yanagida, Phys.
Lett. {\bf B498} 189 (2001); J. Sato and K. Tobe, Phys. Rev. {\bf
D63} 116010 (2001); S. Lavignac, I. Masina and C. A. Savoy,
arXiv:hep-ph/0106245; T.\ Moroi, JHEP.\ {\bf 0003} 019 (2000),
  [arXiv:hep-ph/0002208];  N.\ Akama, Y.\ Kiyo, S.\ Komine and T.\ Moroi,
Phys.\ Rev.\ {\bf D64} 095012 (2001)
  [arXiv:hep-ph/0104263];
T.\ Moroi, Phys.\ Lett.\ {\bf B493}, 366 (2000) [arXiv:
hep-ph/0007328].
\bibitem{lfv3} J. Hisano, T. Moroi, K. Tobe and M. Yamaguchi,
Phys. Rev. D{\bf 53}(1996)2442[hep-ph/9510309]; J. Hisano and D. Nomura,
Phys. Rev. D{\bf 59}(1999)116005[hep-ph/9810479].
\bibitem{cmm}D. Chang, A. Masiero and H. Murayama, hep-ph/0205111.
\bibitem{mvv} A. Masiero, S. K. Vempati and O. Vives,
Nucl. Phys. B{\bf 649}(2003)189[hep-ph/0209303].
\bibitem{bsv}B. Bajc, G. Senjanovi$\acute{c}$ and F. Vissani,
hep-ph/0210207; H.S. Goh, R.N. Mohapatra and S.-P. Ng,
hep-ph/0303055.
\bibitem{hs}J. Hisano and Y. Shimizu, hep-ph/0303071.
\bibitem{lfv} F.~Borzumati and A.~Masiero,  Phys. Rev. Lett. {\bf
57}(1986)961.
\bibitem{lfv2} R. Barbieri and L.J. Hall, Phys. Lett. B{\bf 338}(1994)212;
R. Barbieri, L.J. Hall and A. Strumia, Nucl. Phys. B{\bf 445}(1995)219.
\bibitem{hkr} L.J. Hall, V.A. Kostelecky and S. Raby, Nucl. Phys. B{\bf 267}
(1986)415.
\bibitem{eln}J. Ellis, J.L. Lopez and D.V. Nanopoulos, Phys. Lett.
{\bf B245} (1990) 375; A. Font, L. Ib$\acute{a}\grave{n}$ez and F.
Quevedo, Nucl. Phys. {\bf B345} (1990) 389.
\bibitem{AFDVN}
A.~E.~Faraggi, D.~V.~Nanopoulos and K.~j.~Yuan,
Nucl.\ Phys.\ B {\bf 335}, 347 (1990);
A.~E.~Faraggi, Nucl.\ Phys.\ B {\bf 387}, 239 (1992),
and references therein.
\bibitem{IAGLJR}
I.~Antoniadis, G.~K.~Leontaris and J.~Rizos,
Phys.\ Lett.\ B {\bf 245}, 161 (1990).
\bibitem{JLLDVN}
I.~Antoniadis, J.~R.~Ellis, J.~S.~Hagelin and D.~V.~Nanopoulos,
Phys.\ Lett.\ B {\bf 231}, 65 (1989);
J.~L.~Lopez, D.~V.~Nanopoulos and K.~j.~Yuan,
Nucl.\ Phys.\ B {\bf 399}, 654 (1993), and
 references therein.
\bibitem{fla}
M. Duncan, Nucl. Phys. B{\bf 221}(1983)285;
J. Donoghue, H. Nilles and D. Wyler, Phys. Lett. B{\bf 128}(1983)55;
A. Bouquet, J. Kaplan and C. Savoy, Phys. Lett. B{\bf 148}(1984)69.
\bibitem{smbarr} S. M. Barr,
Phys.\ Lett.\ B {\bf 112}, 219 (1982);
see also
A.~De Rujula, H.~Georgi and S.~L.~Glashow, Phys.\ Rev.\ Lett.\  {\bf 45}, 413 (1980).
\bibitem{dimitri}
J.~P.~Derendinger, J.~E.~Kim and D.~V.~Nanopoulos,
Phys.\ Lett.\ B {\bf 139}, 170 (1984);
I.~Antoniadis, J.~R.~Ellis, J.~S.~Hagelin and D.~V.~Nanopoulos,
Phys.\ Lett.\ B {\bf 194}, 231 (1987).
\bibitem{Fusaoka:1998vc}
H.~Fusaoka and Y.~Koide, Phys.\ Rev.\ D {\bf 57}, 3986 (1998).
\bibitem{rgeff}
 P.H. Chankowski , W. Krolikowski and S. Pokorski,
Phys.Lett. B{\bf 473}(2000)109[hep-ph/9910231] and references therein.
\bibitem{ggms} F. Gabbiani, E. Garieli, A. Masiero and L. Silvestrini,
Nucl.Phys. B477(1996)321[hep-ph/9604387].
\bibitem{mpr} M. Misiak, S. Pokorski and J. Rosiek,
in Heavy Flavours II, eds. A.J. Buras and M. Lindner,
Advanced Series on Directions in High Energy Physics,
World Scientific 1997[hep-ph/9703442].
\bibitem{bcrs} A.J. Buras, P.H. Chankowski, J. Rosiek
and L. Slawianowska, hep-ph/0210145.
\bibitem{gim}S. L. Glashow, J. Iliopoulos, and L. Maiani,
Phys. Rev. D{\bf 2}(1970)1285.
\bibitem{pdg} K. Hagiwara et.al, Review of the Particle Physics,
Phys.RevD{\bf 66}(2002)010001.
\bibitem{cckss} D. Chang, W-F. Chang, W-Y. Keung, N. Sinha and R. Sinha,
Phys. Rev. D{\bf 65}(2002)055010[hep-ph/0109151].
\bibitem{cdm}J. Edsj$\ddot{o}$, hep-ph/0301106; H. Baer and C. Balazs,
hep-ph/0303114; A.B. Lahanas and D.V. Nanopoulos, hep-ph/0303130.
\bibitem{hmty}J. Hisano, T. Moroi, K. Tobe and M. Yamaguchi,
Phys.Lett. B{\bf 391}(1997)341[hep-ph/9605296]; (E)ibid. B{\bf 397}(1997)357.
\bibitem{smr1}B. Mele, S. Petrarca and A. Soddu, Phys. Lett. {\bf
B435} (1998) 401; G. Eilam, J.L. Hewett and A. Soni, Phys. Rev.
{\bf D44} (1991) 1473, Erratum: Phys. Rev. {\bf D59} (1998)
039901.
\bibitem{smr2}W.S. Hou, Phys. Lett. {\bf B296} (1992) 179; K. Agashe
and M. Grzaesser, Phys. Rev. {\bf D54} (1996) 4445; M. Hosch, K.
Whisnant and B.L. Young, Phys. Rev. {\bf D56} (1997) 5725.
\bibitem{gs}J. Guasch and J. Sol$\acute{a}$, hep-ph/9906268. For
earier liturature, C.S. Li, R.J. Oakes, J.M. Yang, Phys. Rev. {\bf
D49} (1994) 293, Erratum: ibid. {\bf D56} (1997) 3156; J.M. Yang
and C.S. Li, Phys. Rev. {\bf D49} (1994) 3412; G. Couture, C.
Hamzaoui, H. K$\ddot{o}$nig, Phys. Rev. {\bf D52} (1995) 1713; G.
Couture, M. Frank, H. K$\ddot{o}$nig, Phys. Rev. {\bf D56} (1997)
4213; J.L. Lopez, D.V. Nanopoulos, R. Rangarajan, Phys. Rev. {\bf
D56} (1997) 3100; G.M. de Divitiis, R. Petronzio, L. Silverstini,
Nucl. Phys. {\bf B504} (1997) 45; J.M. Yang, B. Young and X. Zhang,
Phys. Rev. {\bf D58} (1998) 055001; G. Eilam et al., Phys. Lett.
{\bf B510} (2001) 227.
\bibitem{am} Proceedings of the Workshop on Standard Model
Physics (and More) at the LHC, ed. G. Altarelli and M.L. Mangono,
CERN 2000-004.
\bibitem{cxy}J. Cao, Z. Xiong and J.-M. Yang, hep-ph/0208035;
C.-S. Huang, X.-H. Wu and S.-H. Zhu, Phys. Lett. {\bf B452} (1999)
143; T. Han, J. Hewett, hep-ph/9811237; U. Mahanta, A. Ghosal,
Phys. Rev. {\bf D57} (1998) 1735; Y. Koide, hep-ph/9701261; D.
Atwood, L. Reina, A. Soni, Phys. Rev. {\bf D53} (1996) 1199. For
the earier liturature, see, for example, the references in Phys.
Lett. {\bf B452} (1999) 143 and hep-ph/0208035.
\bibitem{flacp} S. Dimopoulos, L.J. Hall, Phys. Lett. B{\bf 344}(1995)185;
R. Barbieri, L.J. Hall and A. Strumia,
Nucl. Phys. B{\bf 449}(1995)437; Phys. Lett. B{\bf 369}(1996)283.
\bibitem{ach} N. Arkani-Hamed, Hsin-Chia Cheng and L.J. Hall,
Phys. Rev. D{\bf 53}(1996)413.
%\bibitem{mass}J.L. Lopez and D.V. Nanopoulos, Phys. Lett. {\bf B}
%(1991) [CTP-TAMU-27/91].
%\bibitem{emb}S. Ranfone and J.W.F. Valle, Phys. Lett. {\bf B386} (1996) 151.
\end{thebibliography}
\end{document}